\documentclass[a4paper,11pt]{article}
\pdfoutput=1 

\usepackage{jcappub} 
\usepackage{graphicx,color}
\usepackage[dvipsnames]{xcolor}
\usepackage{amsmath}
\usepackage{booktabs}
\usepackage{hyperref}
\usepackage[normalem]{ulem}
\usepackage[T1]{fontenc} 
\usepackage{capt-of}
\usepackage{multirow}

\newcommand{\sigv}{\langle \sigma v \rangle}
\newcommand{\Jcal}{\mathcal{J}}
\newcommand{\mDM}{m_\mathrm{DM}}
\newcommand{\deriv}[2]{\frac{\mathrm{d} #1}{\mathrm{d} #2}}
\newcommand{\bb}{b \overline{b}}

\title{Searching for dark matter subhalos in the \emph{Fermi}-LAT catalog with Bayesian neural networks}

\author[a,b]{Anja Butter,}
\author[c]{Michael Kr\"amer,}
\author[c, d, 1]{Silvia Manconi \note{Corresponding authors}}
\author[c,1]{and Kathrin Nippel}

\affiliation[a]{LPNHE, Sorbonne Universit\'e, Universit\'e Paris Cit\'e, CNRS/IN2P3, Paris, France}
\affiliation[b]{Institut f\"ur Theoretische Physik, Universit\"at Heidelberg, Germany }
\affiliation[c]{Institute for Theoretical Particle Physics and Cosmology, RWTH Aachen University, D-52056 Aachen, Germany}
\affiliation[d]{Laboratoire d'Annecy-le-Vieux de
Physique Théorique (LAPTh), CNRS, USMB, F-74940 Annecy, France}

\emailAdd{silvia.manconi@lapth.cnrs.fr, nippel@physik.rwth-aachen.de}
\arxivnumber{TTK-23-06, LAPTH-013/23}

\abstract{
About a third of the $\gamma$-ray sources detected by the Fermi Large Area Telescope (\emph{Fermi}-LAT) remain unidentified, and some of these could be exotic objects such as dark matter subhalos. We present a search for these sources using Bayesian neural network classification methods applied to the latest 4FGL-DR3 \emph{Fermi}-LAT catalog. We first simulate the $\gamma$-ray properties of dark matter subhalos using models from N-body simulations and semi-analytical approaches to the subhalo distribution. We then assess the detectability of this sample in the 4FGL-DR3 catalog using the \emph{Fermi}-LAT analysis tools. We train our Bayesian neural network to identify candidate dark matter subhalos among the unidentified sources in the 4FGL-DR3 catalog. Our results allow us to derive conservative bounds on the dark matter annihilation cross section by excluding unidentified sources classified as astrophysical-like by our networks. We estimate the number of candidate dark matter subhalos for different dark matter masses and provide a publicly available list for further investigation. Our bounds on the dark matter annihilation cross section are comparable to previous results and become particularly competitive at high dark matter masses.}

\begin{document}
\maketitle
\flushbottom


\section{Introduction}\label{sec:intro}   
According to cosmological simulations, dark matter structures form hierarchically, with dark matter halos containing a large number of smaller substructures known as subhalos (see e.g. \cite{Springel:2008cc, Gao:2004au,Madau:2008fr,Diemand:2008in,Vogelsberger:2014dza,Vogelsberger:2014kha}). A significant fraction of these subhalos are invisible at optical wavelengths because they lack baryonic matter. However, dark matter subhalos in the Milky Way could be detected as $\gamma$-ray point sources due to the annihilation or decay of dark matter particles into Standard Model final states.
The Fermi Large Area Telescope (\emph{Fermi}-LAT) \cite{Fermi-LAT:2019yla, 4FGL-DR3} has detected over six thousand $\gamma$-ray point sources, most of which are blazars and pulsars. However, more than two thousand sources remain without clear identification or counterparts at other wavelengths. Some of these unidentified (UNID) sources could potentially be attributed to dark matter annihilation or decay in subhalos. 

Signatures of dark subhalo populations have been searched for within individual $\gamma$-ray sources in \emph{Fermi}-LAT catalogs~\cite{2012A&A...538A..93Z,Buckley:2010vg,Bertoni:2015mla,Schoonenberg:2016aml,Hooper:2016cld,PhysRevD.96.063009,Coronado-Blazquez:2019pny,Coronado-Blazquez:2019puc}, or as a collective contribution to diffuse emissions \cite{Bringmann:2013ruh,Fornasa:2015qua}, detectable by modelling their intensity \cite{DiMauro:2015tfa,Fermi-LAT:2015qzw,Hutten:2017cyu, Huetten_2019, Calore:2019lks} or by their angular fluctuations \cite{Calore:2014hna,Fornasa:2016ohl}. Subhalos are expected to appear in \emph{Fermi}-LAT surveys as steady sources of $\gamma$ rays, predominantly  point-like with possibly few extended nearby representatives \cite{DiMauro:2020uos,Coronado-Blazquez:2021amj}. Furthermore, subhalos should have no counterpart emission at other wavelengths, and their $\gamma$-ray emission is expected to be signal dominated, with little background from other astrophysical mechanisms.

Machine learning classifiers have been used in recent years to study UNID sources \cite{Mirabal:2016huj, Saz_Parkinson_2016, Luo_2020, Finke:2020nrx, Bhat:2021wtb, Panes:2021zig, Butter:2021mwl, Sahakyan:2022uvr} and to search for potential dark subhalos. Using XGBoost classifiers on sources in the third \emph{Fermi}-LAT catalog, ref.~\cite{Mirabal:2016huj} searched for Galactic and dark matter subhalo candidates with pulsar-like spectra outside the Galactic plane, and further analysed this list with multiwavelength observations. The work presented in ref.~\cite{Mirabal:2021ayb} explored UNID sources in the fourth \emph{Fermi}-LAT catalog, exploiting the similarity between pulsar and dark subhalo spectra for certain dark matter models, and investigated the spatial distribution of the candidates by comparing it to the expected structure of the GD-1 perturbator. In the recent ref.~\cite{Gammaldi:2022wwz}, the authors base their classification on the spectral features expected from a generic dark matter spectrum \cite{Coronado-Blazquez:2019pny}, and artificially build systematic features sampling from the real observed sources to improve the performance of the network.
Constraints on dark matter properties based on subhalo searches usually formulate their results as a function of the possible number of candidates among UNID sources \cite{PhysRevD.96.063009,Coronado-Blazquez:2021amj}.  
The results of machine learning classification on \emph{Fermi}-LAT sources were used as a further filtering step to select subhalo candidates in ref.~\cite{Coronado-Blazquez:2019puc}. However, all these results were based on machine learning classifiers that were not trained on realistic subhalo simulations, 
but only using their expected similarities to observed astrophysical sources. 

In this study, we train a Bayesian neural network to reliably differentiate between astrophysical sources  and sources that may originate from dark matter annihilation in subhalos. We study a simple and generic Majorana dark matter model with annihilation into bottom quark states. The approach we present is, however, not limited to this specific model. 

We model the dark matter subhalo population using the \texttt{CLUMPY} code based on dark matter-only cosmological simulations. We calculate the expected flux of $\gamma$ rays in the \emph{Fermi}-LAT energy range for each subhalo using different dark matter spectral models. To compare these modelled subhalos with the observed $\gamma$-ray spectra, we simulate the measurement of $\gamma$ rays with \emph{Fermi}-LAT using the \textsc{fermipy} tool. Our first new result is an estimate of the detectability of dark matter subhalos for different dark matter models. Furthermore, we train the Bayesian neural network on the $\gamma$-ray spectra of the detectable subhalos, and use the network to identify dark matter subhalo candidates among the \emph{Fermi}-LAT UNID sources. The Bayesian neural network classifier also allows us to set conservative limits on the dark matter annihilation cross section, which are particularly competitive at large dark matter masses. We also demonstrate the importance of separately evaluating the number of subhalo candidates in \emph{Fermi}-LAT catalogs for each dark matter model.

Our work extends previous work in several directions. First, we train our networks on realistic dark matter subhalo simulations, taking into account the specific spectral properties produced by different dark matter masses. Second, we base the classification on the measured flux as a function of energy, instead of using human-created features,
i.e. we use the full information contained in the measured energy spectra.  
Finally, we use Bayesian neural networks, which allow us to account for uncertainties in the network weights and make robust predictions about subhalo-like sources. 

In section~\ref{sec:SHphys} we present the physical properties and assumptions of dark matter subhalos, including the expected $\gamma$-ray spectra emitted by subhalos, and their simulation. We then outline the detection of these spectra in the \emph{Fermi}-LAT data as a second simulation step in section~\ref{sec:data_sim}. The results of these simulations are discussed in section~\ref{sec:results_det}, where we study the detectable dark matter subhalos for different dark matter models. Our machine learning approach is introduced in section~\ref{sec:ml_class}, where we present our results on the classification of \emph{Fermi}-LAT unidentified sources. Finally, using these results, we present our limits on the dark matter annihilation cross section for our dark matter model in section~\ref{sec:limits}. We conclude in section~\ref{sec:conclusions} and present further technical details of our analysis in the appendices.

%
\section{Dark matter subhalos as $\gamma$-ray sources}
\label{sec:SHphys}
Cosmological simulations predict that dark matter is clustered hierarchically in galactic halos \cite{Gao:2004au,Madau:2008fr,Diemand:2008in,Vogelsberger:2014dza,Vogelsberger:2014kha}. Larger structures are the product of accretion and the merging of smaller structures. As a result of this structure formation model, a large number of substructures, called subhalos, are predicted to populate dark matter halos such as the Milky Way. The properties of these substructures, such as their density profile and their spatial evolution, depend on various model assumptions, see ref.~\cite{Zavala:2019gpq} for a comprehensive review. 

A broad mass spectrum for dark matter substructures is observed in simulations, with an abundance well described by a power law $dN/dM \propto M^{-\alpha_M}$, with a slope $\alpha_M\sim 1.9$ over many orders of magnitude \cite{Zavala:2019gpq}. The role of low-mass subhalos is currently under debate, and depends on the extrapolation of the subhalo mass function down to masses below the resolution of the simulations, $M\lesssim 10^{4-6}$~M$_{\odot}$ \cite{Coronado-Blazquez:2019puc,Aguirre-Santaella:2022kkm}. At even lower masses of around $m_{\rm min} \approx 10^{-(4-10)}M_{\odot}$, a sharp cutoff in the mass spectrum is expected, which can be related to the kinetic decoupling of dark matter particles in the early universe \cite{Berezinsky:2003vn,Bringmann:2006mu}.

The more massive substructures in dark matter halos form dwarf galaxies, which contain subdominant baryonic matter in gas and stars and which provide strong constraints on dark matter properties \cite{Alvarez:2020cmw,Hess:2021cdp}. The Milky Way's dwarf galaxies are thought to be representative of a larger population of satellite substructures which may lack any baryonic counterpart in stars or gas, and are thus completely 'dark' to e.g.\ optical surveys \cite{DES:2020fxi}.

Subhalos are thus structures predominantly composed of dark matter. If dark matter particles are weakly interacting massive particles (WIMPs) that can self-annihilate into Standard Model particles, then individual subhalos could shine as sources of emission in the sky, potentially offering a signal for dark matter. According to the WIMP paradigm, natural dark matter candidates with masses in the GeV-TeV range could produce $\gamma$ rays through annihilation into hadronic or leptonic final states.

The differential $\gamma$-ray flux, $\phi_\gamma$, from the annihilation of (Majorana) WIMP particles from a dark matter subhalo is computed as:
\begin{equation}
   \phi_\gamma : = \deriv{\Phi_{\gamma}}{E}(E, \Delta \Omega) = \frac{\sigv}{8  \pi \mDM^2} \; \mathcal{J}(\Delta \Omega)  \; \deriv{N_{\gamma}^i}{E} (E) \, ,
   \label{eq:flux}
\end{equation}
where  $\mDM$ is the dark matter mass, $\sigv$ is the thermally averaged annihilation cross section,  $\Jcal$ is the so-called $\Jcal$-factor, which is the integral along the line of sight of the subhalo dark matter density over the solid angle $\Delta \Omega$,  and $\mbox{d}N^i_\gamma(E)/\mbox{d}E$ is the energy spectrum of $\gamma$ rays produced by dark matter annihilation in a given annihilation channel $i$. Note that the first two factors in eq.~\eqref{eq:flux} specify the flux normalisation, while the energy dependence is only determined by $\mbox{d}N^i_\gamma(E)/\mbox{d}E$. How we model the $\Jcal$-factor and the energy spectrum in eq.~\eqref{eq:flux} is described in more detail in the following sections.

\subsection{Galactic subhalo population model}

Our focus is on the population of dark substructures in the Milky Way, which can be modelled by studying the evolution of dark matter halos similar to our Galaxy through either fully numerical cosmological simulations or simplified analytical models~\cite{Stadel:2008pn,Diemand:2007qr,Madau:2008fr,Garrison-Kimmel:2013eoa,Hiroshima:2022khy}. While the signals from extragalactic substructures have also been studied~\cite{Ullio:2002pj,Ando:2013ff,Fermi-LAT:2015qzw,Hutten:2017cyu,Bartlett:2022ztj}, we do not consider them in this analysis as we adopt a conservative approach.

The population of substructures within the host dark matter halo is characterised by their abundance, mass, spatial distribution, and the internal structure of each subhalo.  
These properties are determined by the initial cosmological conditions and various dynamical processes, including dynamical friction, tidal stripping and heating \cite{Zavala:2019gpq}. However, modelling these phenomena is challenging and introduces various levels of uncertainty, leading to uncertainties in the properties of the Galactic subhalo population and the resulting annihilation signal~\cite{PhysRevD.96.063009,Huetten_2019,Calore:2019lks,Stref:2019wjv}.

We adopt the \textit{Aquarius} dark matter-only N-body simulation \cite{Springel:2008cc} as our primary subhalo model, which has also been used in previous studies \cite{Huetten_2019,Calore:2019lks} and is referred to as \texttt{DM-only}. 
Compared to other models including e.g.\ baryonic effects \cite{Calore:2019lks}, this model is the most optimistic in terms of the number of predicted bright subhalos. 
Furthermore, observations suggest that the dark matter profile of halos and subhalos could possibly deviate from what inferred from N-body simulations, see e.g. refs.~\cite{Alfaro_2023, Karukes2016}. 
We choose the \texttt{DM-only} model for its simplicity and to facilitate comparison with previous work. While more refined models that account for additional physical effects are available, we leave their exploration to future investigations.

The spatial distribution of the subhalos within the main halo is described by an Einasto profile \cite{Springel:2008cc} ($\alpha_E=0.68$, $r_{-2}=199$~kpc), while the dark matter density of each subhalo follows a Navarro-Frenk-White \cite{Navarro:1995iw} profile. 
The mass function is modelled as a power law with $\alpha_M=1.9$, and the mass-concentration relation follows that of ref.~\cite{Moline:2016pbm}. 
We refer to \cite{Calore:2019lks,Huetten_2019} for any further details on the model and its parameters.

We use the \texttt{CLUMPY} code \cite{clumpyv1,clumpyv2,clumpyv3} to simulate subhalos according to the \texttt{DM-only} model. \texttt{CLUMPY} is a publicly available code written in C++ which computes the properties of dark matter structures semi-analytically. In addition, \texttt{CLUMPY}  provides $\gamma$-ray and neutrino fluxes from dark matter annihilation or decay for our Galaxy and for the mean extragalactic contribution.

We compute the $\Jcal$-factor for each subhalo by integrating the square of the dark matter density $\rho_{\rm DM}$ along the line of sight (l.o.s.):
\begin{equation}\label{eq:Jsub}
    \mathcal{J}_{\rm sub}(\Delta \Omega) = \int_0 ^{\Delta \Omega} d\Omega \int_{\rm l.o.s.} \rho_{\rm DM}^2 \; dl \;, 
\end{equation}
where we also integrate over the solid angle $\Delta \Omega=2\pi (1-\cos \theta_{\rm int})$ and treat each subhalo as a point-like source by setting the integration angle $\theta_{\rm int}=0.5$~deg.  This choice is in line with our simulation strategy for point-source subhalos (see next section) and is close to the angular resolution of \emph{Fermi}-LAT at energies of a few GeV \cite{WinNT}. Larger $\theta_{\rm int}$ values (e.g. up to the profile scale radius) may lead to an overestimation of the $\Jcal$-factor and, consequently, overly stringent dark matter limits \cite{Calore:2019lks,DiMauro:2020uos}. 

We only consider subhalos with $\mathcal{J}_{\rm sub}(<0.5\,\rm deg)$ values greater than $10^{17}$~GeV$^2$cm$^{-5}$, as subhalos with smaller values are unlikely to produce detectable $\gamma$-ray emission in current \emph{Fermi}-LAT catalogs for the dark matter models we consider, see also \cite{Calore:2019lks}. We have tested that changing the integration angle to $\theta_{\text{int}}=0.1$~deg does not bring a sizeable effect in the values or numbers of subhalos for $\Jcal_{\rm sub} > 10^{17} \, \mathrm{GeV}^2 \, \mathrm{cm}^{-5}$.

\begin{figure}[t]
	\center
	\includegraphics[width=0.9\textwidth]{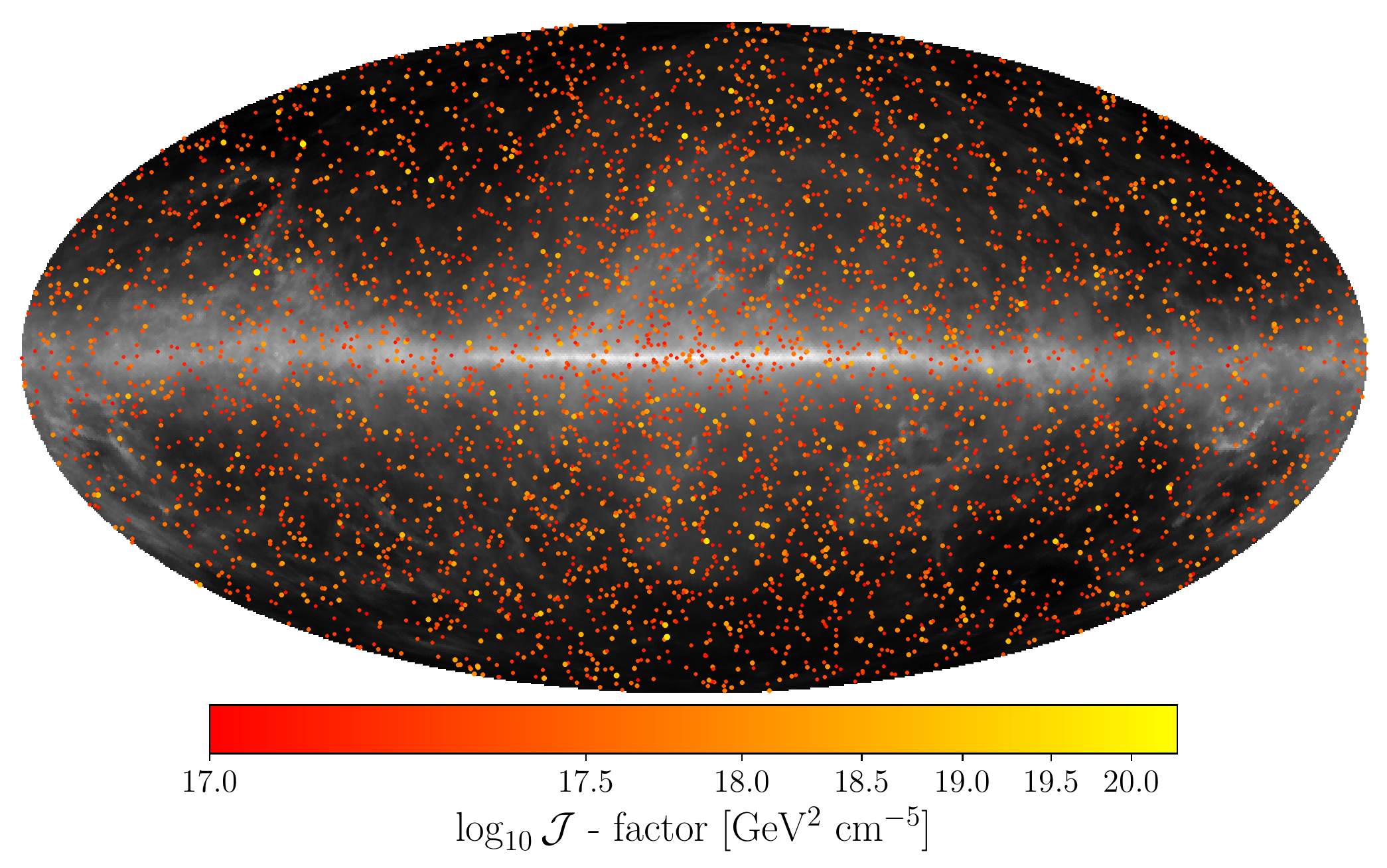}
		\caption{Skymap showing the position and the 
		$\Jcal$-factor value for a population of Galactic subhalos obtained using the \texttt{DM-only} model simulated using  \texttt{CLUMPY}. The colour scale and the marker size illustrate the value of the $\Jcal$-factor; we cut at $\Jcal < 10^{17} \, \mathrm{GeV}^2 \, \mathrm{cm}^{-5} $. The diffuse emission coming from cosmic ray interactions is overlaid using light grey colours, highlighting the bright background emission along the Galactic plane.}
		\label{fig:Jfac_skymap}
\end{figure}

In figure~\ref{fig:Jfac_skymap}, we present a Mollweide projected skymap that illustrates the distribution of subhalos and their corresponding $\Jcal$-factor values for a single \texttt{CLUMPY} realisation, obtained following the \texttt{DM-only} model as described earlier. Each subhalo with $\Jcal_{\rm sub} > 10^{17} \, \mathrm{GeV}^2 \, \mathrm{cm}^{-5}$ is marked by a dot, with the size and colour of the dot representing the corresponding $\Jcal_{\rm sub}$ value as computed using eq.~\eqref{eq:Jsub}. In the map, we have overlaid the background $\gamma$-ray emission that arises from cosmic ray interactions in light grey. We use the background model  released and recommended by the \emph{Fermi}-LAT collaboration, which has been optimised for analysing $\gamma$-ray sources in the \emph{Fermi}-LAT energy range.

In realistic \emph{Fermi}-LAT simulations, the detectability of dark subhalos is expected to depend on their position in the sky due to the higher background coming from the interstellar emission at low Galactic latitudes ($|b|<20$~degrees), see e.g.\ \cite{PhysRevD.96.063009}. To account for this, we calculate the emission from each dark subhalo at its simulated position in the sky, taking into consideration the interstellar background and potential contamination from nearby point sources. Since the \texttt{CLUMPY} results are independent of the assumed dark matter mass and annihilation spectrum, we use the same statistical realisation of subhalos with their $\Jcal$-factor and sky positions (galactic longitude, $l$, and latitude, $b$) for each benchmark model discussed in the following sections.

State-of-the-art N-body simulations can now account for the effect of the baryonic potential in the dynamics of dark matter clustering, including the stellar and gas disk, the bulge of galaxies, as well as magnetic fields and feedbacks from supernovae and active galactic nuclei \cite{Dubois:2014lxa,Vogelsberger:2014dza,Schaye:2014tpa,Pillepich:2017jle}.
This affects the resulting distribution of substructures on Galactic scales with respect to the \texttt{DM-only} case, and thus the dark matter annihilation signal, as quantified e.g.\ in refs.~\cite{Calore:2019lks,Huetten_2019}. 
Accounting for these effects is beyond the scope of the present paper. However, we note that extending our approach to other dark matter subhalo models would be straightforward and will be considered in future work. Examining eq.~\eqref{eq:flux}, we note that different subhalo models would, to a good approximation, only affect the distribution of the resulting $\mathcal{J_{\rm sub}}$, and not the annihilation spectra. We therefore expect that the algorithms presented in section~\ref{sec:ml_class} to distinguish between dark matter subhalos and astrophysical source spectra would work similarly, and only the total number of detectable subhalos would change, similar to what was observed in \cite{Calore:2019lks}. 

\subsection{ $\gamma$-ray spectra from dark matter annihilation}
\label{sec:gamma_spectra}
The $\gamma$-ray flux resulting from the annihilation of WIMP particles, eq.\eqref{eq:flux}, depends on the annihilation cross section $\sigv$, the dark matter mass $\mDM$, and the annihilation channel~$i$. We focus on the hadronic annihilation channel $i= b \bar{b}$ with a branching ratio of one. This channel produces energy spectra similar to those of other astrophysical sources, such as pulsars, and represents the most challenging setup for our machine learning classifiers. Our results can be extended to account for other annihilation channels and the branching ratio structure of specific particle physics models, see e.g.\ \cite{Armand:2022sjf}.

For dark matter masses $\mDM$ above about 10~GeV, the annihilation cross section $\sigv$ required to account for the dark matter relic abundance is nearly independent of the dark matter mass~\cite{Steigman:2012nb}. Thus, to maintain a model-agnostic approach, we treat $\sigv$ and $\mDM$ as independent parameters and derive limits on $\sigv$ as a function of $\mDM$.

\begin{figure}[t]
	\center
	\includegraphics[width=0.75\textwidth]{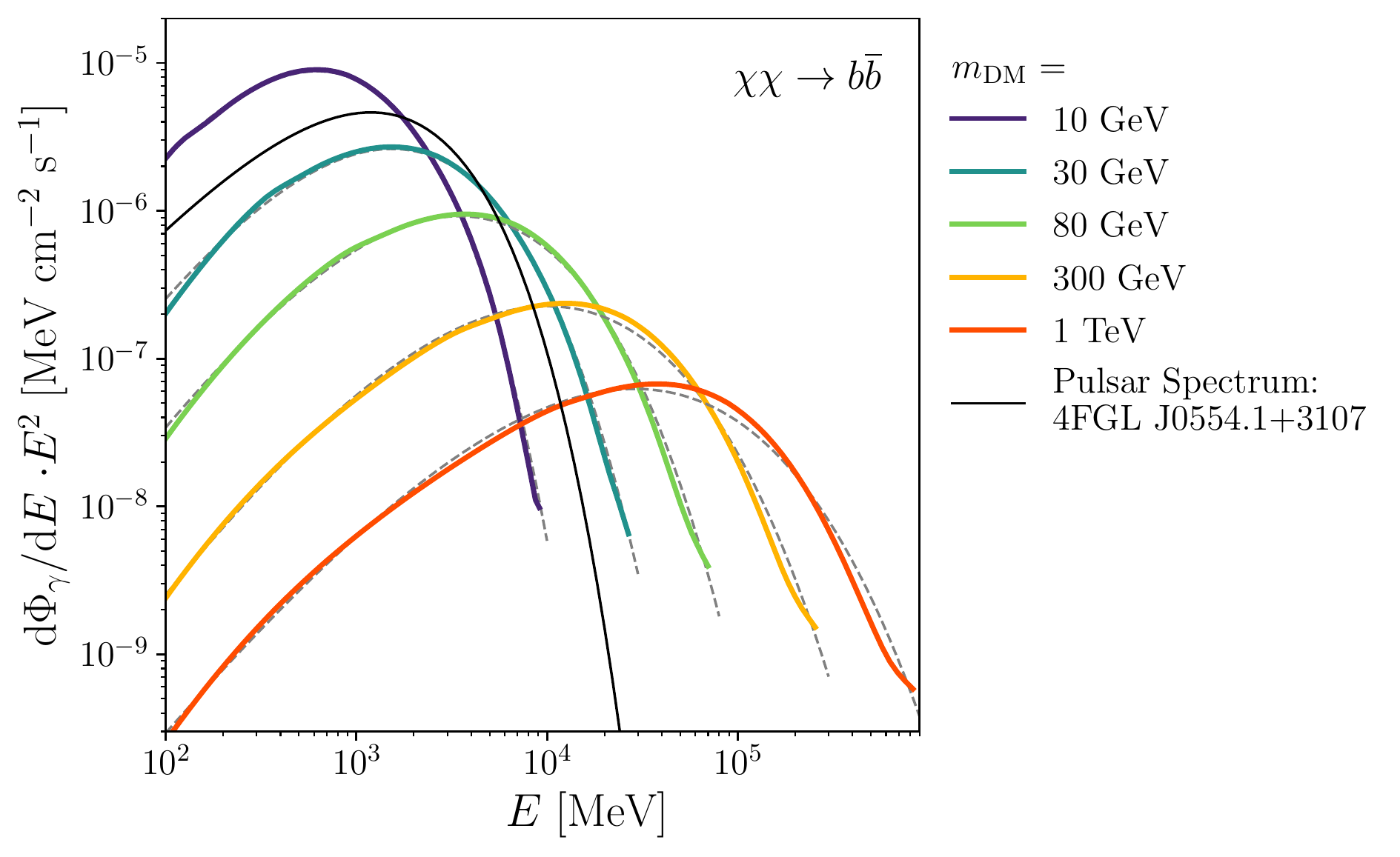}
		\caption{Differential $\gamma$-ray flux from dark matter annihilation into $b \bar{b}$.  We assume a dark matter subhalo with an illustrative normalisation factor of $\sigv \cdot \Jcal = 10^{-5}$. Five different dark matter masses between 10~GeV (blue line) and 1~TeV (red line) are displayed.  The dashed lines show the fit functions we use to simulate the measurement of the $\gamma$-ray flux with \textsc{fermipy}. An example pulsar spectrum from the 4FGL-DR3 catalog is included for comparison (solid black line).}
		\label{fig:dNdE_PPPC}
\end{figure}

We use the $\gamma$-ray energy spectra ${\mathrm{d}N_{\gamma}^i(E)}/{\mathrm{d}E}$ as provided in \cite{Cirelli:2010xx},  and include only prompt emission, as secondary emission is negligible for the parameter space explored in our analysis. Figure~\ref{fig:dNdE_PPPC} displays the differential $\gamma$-ray flux from dark matter annihilation into $b\bar{b}$ final states. We use eq.~\eqref{eq:flux} and assume an example subhalo with a normalisation factor of $\sigv \times \Jcal = 10^{-5}$. The plot shows the energy spectra for various dark matter masses $\mDM$ between 10~GeV and 1~TeV. For this mass range, the annihilation flux peaks within the \emph{Fermi}-LAT sensitivity range. We overlay an example pulsar spectrum taken from the source 4FGLJ0554.1+3107. Comparing the pulsar and dark matter spectra, we observe that their shapes become very similar for the $b\bar{b}$-channel and dark matter masses around a few tens of GeV.

Following this observation, some studies \cite{Mirabal:2012em,Coronado-Blazquez:2019puc} have used machine learning classifiers that were trained to identify pulsars among \emph{Fermi}-LAT unidentified sources to estimate the number of dark matter subhalo candidates. However, it is important to note that the similarity between the dark matter and pulsar spectra is limited to a specific range of dark matter masses and specific annihilation channels. Furthermore, it is necessary to include realistic subhalo sources in the training set of the algorithms to ensure consistency when using machine learning classifiers. We will address these points in more detail when we evaluate the accuracy of our classifiers for the simulated subhalo spectra in section~\ref{sec:ml_class}.

As described in the next section, we will simulate the expected $\gamma$-ray flux from each subhalo in the \emph{Fermi}-LAT data by using the  analysis program {\textsc{fermipy}}. However, this analysis program does not support arbitrary input spectra; instead, they must fall within the available functional forms. To accommodate this requirement, we represent the $\gamma$-ray spectrum using a power law with a super-exponential cutoff (PLSuperExpCutoff) within {\textsc{fermipy}}:
\begin{equation}\label{eq:PLSuperExpCutoff}
\deriv{N_{\gamma} }{E} = N_0 \left(\frac{E}{E_0} \right)^\gamma  \exp \left( -  \left(\frac{E}{E_0} \right)^\beta \right) \, ,
\end{equation}
where  the fit parameters are $N_0$, a prefactor with units [MeV$^{-1}$ cm$^{-2}$ s$^{-1}$], the cutoff energy $E_c$ [MeV], and the indices $\gamma$ and $\beta$. The scale $E_0$ is fixed at 1000~MeV. 
It has been shown~\cite{PhysRevD.96.063009} that this fit is suitable to describe the functional form of the $\gamma$-ray energy distribution of annihilating WIMPs. We repeat the fit and find very similar results, with the relative difference between the spectrum and the fit always less than a few percent for energies around the peak of the spectrum. Our fits are shown in figure~\ref{fig:dNdE_PPPC} as dashed lines, for each dark matter mass and channel, for completeness. Further details are given in appendix~\ref{app:spec}.

%
\section{\emph{Fermi}-LAT simulation}
\label{sec:data_sim}
In this section we describe how we simulate the measurement of $\gamma$ rays from dark matter subhalos with \emph{Fermi}-LAT, and how we assess whether a subhalo would be detected in current \emph{Fermi}-LAT catalogs. Our objectives are twofold: (1) to update the detection prospects for point-like dark matter subhalos for the 4FGL-DR3 catalog, and (2) to produce a realistic training set for our neural networks. 
To achieve these objectives, our simulation must produce the following outcomes:

\begin{enumerate}
    \item[i)] The number of detectable subhalos in the 4FGL-DR3 catalog as a function of the annihilation cross section and the dark matter mass; 
    \item[ii)] The energy spectrum that would be detected by Fermi-LAT for each subhalo, taking into account realistic systematic uncertainties from diffuse and point-source backgrounds, the \emph{Fermi}-LAT instrument response function, and using detection pipelines typically employed in \emph{Fermi}-LAT source analysis.
\end{enumerate}  

Since the $\gamma$-ray spectra will be the input for training our neural networks, it is crucial for the simulation of subhalo spectra to reproduce  the statistical and systematic uncertainties of a \emph{Fermi}-LAT measurement as closely as possible. 

Thus, we have extended the simulation strategy, building upon previous work~\cite{PhysRevD.96.063009,Calore:2019lks,DiMauro:2020uos,Coronado-Blazquez:2021amj}, to create realistic training data. In particular we use the latest published data, include the full information from the energy spectra and optimise the fitting procedure to extract these spectra. By processing complete spectra, we can leverage the ability of neural networks to extract all relevant information from low-level features. Our work thus differs from previous studies, such as \cite{Gammaldi:2022wwz}, which used synthetic features to search for dark matter subhalo candidates.

\begin{figure}[t]
	\center
	\includegraphics[width=.65\textwidth]{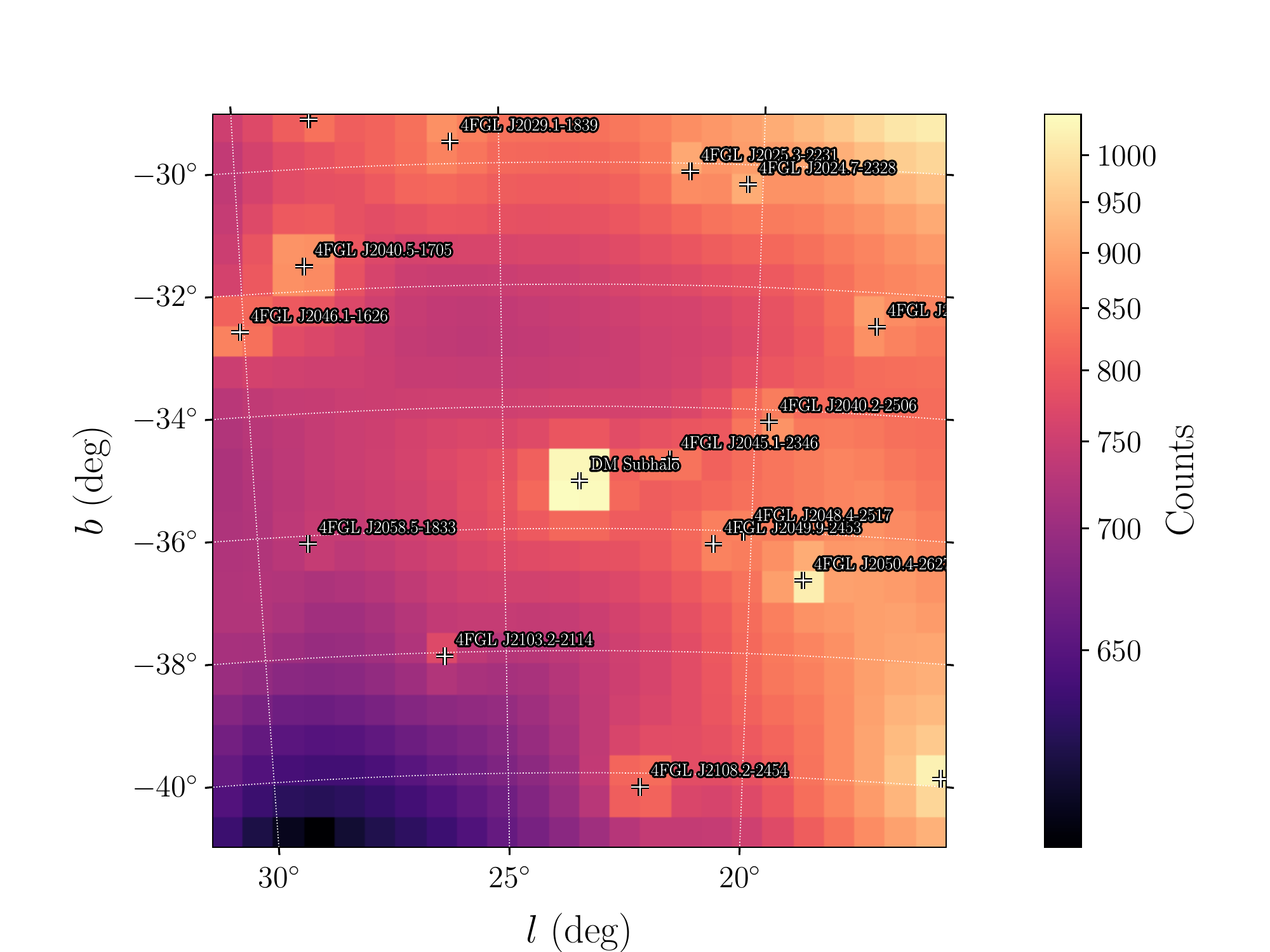}
		\caption{Simulated count map (obtained after the second step as described in section~\ref{sec:data_sim}) for a representative dark matter subhalo located at the centre of the ROI. The position of the sources in the 4FGL-DR3 catalog is also given.} 
		\label{fig:roi_counts}
\end{figure}

We simulate 12 years of \emph{Fermi}-LAT data and apply cut selections compatible with the most recent release of the public \emph{Fermi}-LAT source catalog, the 4FGL-DR3 catalog \cite{4FGL-DR3, Fermi-LAT:2019yla}, see appendix~\ref{app:fermipy_details} and table~\ref{tab:fermipy_stats}. 
The simulations and subsequent data analysis are performed using \textsc{fermipy}, a Python interface to the official analysis tools  developed by the \emph{Fermi}-LAT collaboration\footnote{\url{https://fermipy.readthedocs.io/en/latest/}}. 
For each subhalo, we perform the following four steps: 

\begin{enumerate}
\item \textit{Setup.} For each subhalo generated by the \texttt{CLUMPY} simulation, we define a region of interest (ROI) with dimensions of 12 degrees $\times$ 12 degrees centered at the Galactic longitude and latitude ($l,b$) of the subhalo. This ROI size is similar to previous studies \cite{PhysRevD.96.063009,Coronado-Blazquez:2021amj} and includes photons from the central subhalo and the brightest nearby sources, considering the point spread function at the lowest energy range of our analysis (at 100~MeV, the $68$\% containment angle is approximately 5 degrees \cite{WinNT}). To normalise the flux for each subhalo, we use the $\Jcal$-factor obtained from the \texttt{CLUMPY} simulation and the annihilation cross section of the selected setup, following eq.~\eqref{eq:flux}. 
For each choice of the dark matter mass, we input the $\gamma$-ray spectra parameterised as explained in section~\ref{sec:gamma_spectra}. 
We exclude subhalos with $\Jcal < 10^{17} \, \mathrm{GeV}^2 \, \mathrm{cm}^{-5}$, as their $\gamma$-ray flux is expected to be significantly below the detection threshold \cite{Calore:2019lks}. At this stage, the ROI model includes the astrophysical background, which is composed of diffuse emissions and all the relevant point sources from the 4FGL-DR3 catalog.
\item \textit{Photon event simulation.} We simulate photon events of the full ROI model, which includes diffuse backgrounds, catalog sources, and the dark matter subhalo. Our simulation takes as input the number of counts predicted by the current model of the ROI for 12 years of observation, as well as the instrument response function. Figure~\ref{fig:roi_counts} shows the counts map, i.e., the number of photons integrated in the considered energy range, of the simulated photons for a representative bright subhalo located at intermediate negative latitudes ($l = 23.48^{\circ}, b = -35.22^{\circ}$). We also indicate the position and name of \emph{Fermi}-LAT catalog sources in the ROI. Note that diffuse emissions and nearby 4FGL-DR3 sources contribute significantly to the counts in the ROI at the location of the dark matter subhalo. Therefore, we simulate all these components jointly and optimise their parameters in the next step.  
\item \textit{ROI fit.} We then optimise the model parameters to describe the mock data. This includes simultaneously fitting the parameters of the sources in the model of the ROI and computing their spectral properties and detection significance. We quantify the detection significance using the test statistic (TS) \cite{1996ApJ...461..396M}, defined as:
\begin{equation}\label{eq:ts}
\mbox{TS} = - 2 \log{\frac{\mathcal{L}}{\mathcal{L}_0}},
\end{equation}
where $\mathcal{L}$ and $\mathcal{L}_0$ are the maximum likelihood values with and without the source included in the model of the ROI, respectively. The likelihoods are calculated as the product of the Poisson probabilities of observing a given number of counts, as predicted by the respective model, in each pixel of the ROI.\footnote{\url{https://fermi.gsfc.nasa.gov/ssc/data/analysis/documentation/Cicerone/Cicerone_Likelihood/Likelihood_overview.html}}
\item \textit{Spectral energy distribution.} As a final step, we calculate the spectral energy distribution (SED) of each subhalo by fitting the flux normalisation of the source in each energy bin defined by the 4FGL-DR3 catalog. We adopt the same binning as the 4FGL-DR3 catalog, which has energy bin bounds of $[0.05, 0.1, 0.3, 1, 3, 10, 30, 300, 1000]\,\mathrm{GeV}$, to obtain a consistent data structure with the 4FGL-DR3 detected sources. Thus, the input data for the neural networks is the photon flux in the energy bins defined above. For a significant number of subhalos, the low and high energy bins of the SED are upper limits due to systematic uncertainties (mostly related to background modelling) and statistical uncertainties. Therefore, we also store the test statistics bin-wise for future use.
\end{enumerate}

At the end of our analysis chain, we obtain for each dark matter subhalo: the detection significance (TS) over the full energy range, the detection significance in each energy bin, and the spectral energy distribution that would be measured by \emph{Fermi}-LAT catalogs. The TS values are particularly relevant to the results presented in section~\ref{sec:results_det}, while the spectral energy distributions are crucial for the work detailed in section~\ref{sec:ml_class}. We use both sets of results in section~\ref{sec:limits} to derive limits on the dark matter annihilation cross section. Technical details on the implementation of our analysis pipeline using \textsc{fermipy}\ are provided in appendix~\ref{app:fermipy_details}.

%

%
\section{Detectability of dark matter subhalos with \emph{Fermi}-LAT}\label{sec:results_det}

The detectability of dark matter subhalos in \emph{Fermi}-LAT data depends on several factors, as explored in~\cite{Schoonenberg:2016aml,Bertoni:2015mla, PhysRevD.96.063009,Coronado-Blazquez:2019pny,Coronado-Blazquez:2019puc} and briefly summarised below. We note that the simulation strategy outlined in the previous section allows us to properly account for these effects in our estimate of the number of detectable dark matter subhalos in \emph{Fermi}-LAT data. 

The $\gamma$-ray flux, eq.~\eqref{eq:flux}, depends on the dark matter model, and for WIMP annihilation in particular on the thermally averaged annihilation cross section, the dark matter mass and the annihilation channel. Different combinations of dark matter mass and annihilation channel will produce a $\gamma$-ray flux with a different normalisation and energy spectrum, peaking at different energies, see figure~\ref{fig:dNdE_PPPC}.

The detectability of $\gamma$-ray sources depends on their spectral properties, with harder spectra being easier to detect with the LAT telescope \cite{2015ApJS..218...23A,PhysRevD.96.063009}. Furthermore, the point spread function (PSF) and the acceptance of \emph{Fermi}-LAT depend on energy \cite{WinNT}. Therefore, dark matter spectra that peak at energies where the PSF is smaller 
(68\% containment angle decreases from about 1\,deg at 1\,GeV to 0.1\,deg at 100\,GeV) and the acceptance is larger (from 2.1 to 2.5 m$^2$sr going from 1\,GeV to 100\,GeV) are easier to detect. 

Furthermore, the $\gamma$-ray flux is directly proportional to the $\Jcal$-factor, as defined in eq.~\eqref{eq:Jsub}. The integration angle of the $\Jcal$-factor is set to 0.5 degrees, which is close to the angular resolution of the \emph{Fermi}-LAT instrument at energies in the GeV range. It is important to note that integrating the $\Jcal$-factor up to larger radii can potentially lead to an overestimation of the $\gamma$-ray flux, which could result in non-conservative limits \cite{Calore:2019lks}. 

Finally, the detectability of subhalos depends on their position in the sky. As illustrated in figure~\ref{fig:Jfac_skymap}, at low latitudes a potential subhalo source must appear above a  bright Galactic diffuse emission. This increases both the minimum flux that a subhalo must emit in order to be detected with significance, and the systematic uncertainties in its measured properties. 
For this reason, realistic detectability studies have either considered the location of subhalos at different fixed latitudes, assuming that the longitude dependence of the background is less significant \cite{PhysRevD.96.063009}, or have computed the full LAT sensitivity at each sky position \cite{Coronado-Blazquez:2019puc}. In this work, we take into account the full spatial dependence of the backgrounds. Specifically, we locate each dark matter subhalo at its simulated position in the sky, and consistently simulate the Galactic interstellar emission in the corresponding ROI. This approach ensures that the potential impact of the background on the detectability of subhalos is properly accounted for.

To assess the detectability of dark matter subhalos, we focus on WIMP annihilation into $b\bar{b}$ final states and consider a wide range of annihilation cross sections $\sigv$ and dark matter masses $\mDM$. Our benchmark setup corresponds to $\mDM = 80\, \mathrm{GeV}$ and $\sigv = 10^{-23}\, \mathrm{cm}^3\mathrm{s}^{-1}$. As discussed in section~\ref{sec:ml_class}, the annihilation cross section of the benchmark setup is chosen to provide a sufficient amount of training data for the neural network classifier. 

\subsection{Detection significance}
\begin{figure}[t]
	\center
	\includegraphics[width=\textwidth]{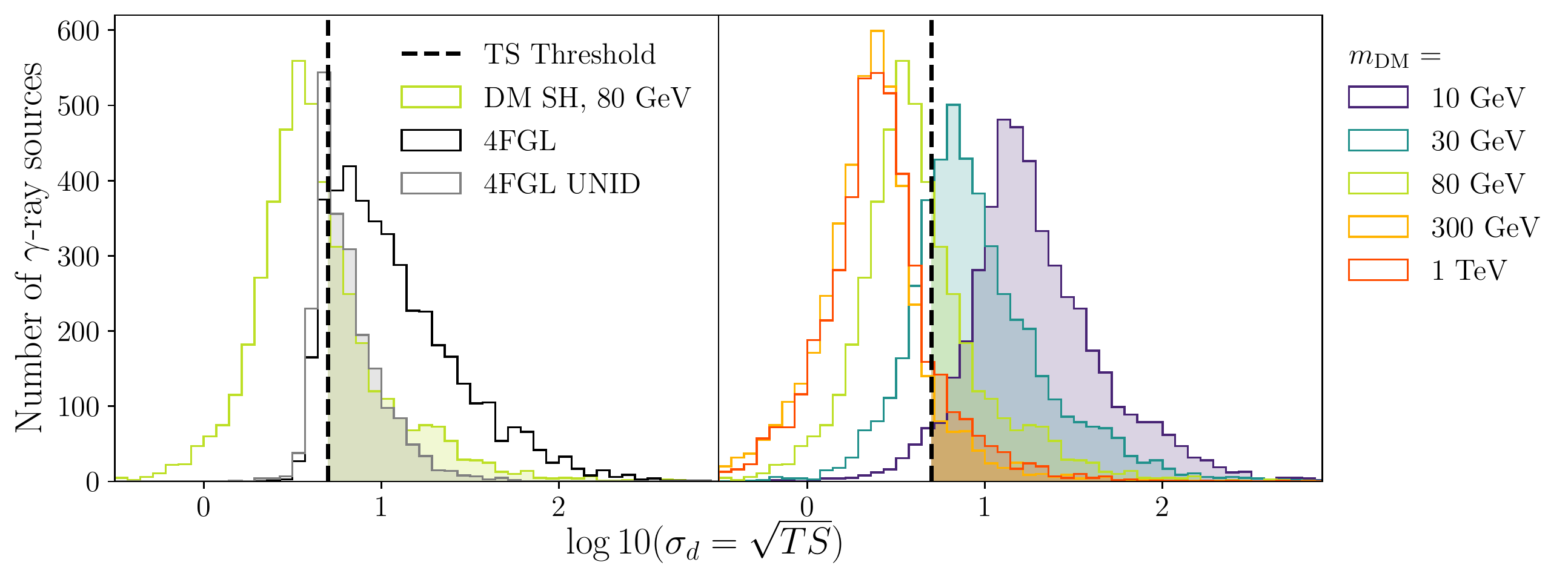}
		\caption{\textit{Left:} Histogram of the detection significance for the dark matter subhalos in our benchmark setup \textit{(yellow)}. All halos measured with ${\rm TS} \geq 25$ \textit{(dashed line)} are added to the catalog of artificial sources. The classified \textit{(black)} and UNID \textit{(grey)} sources in the 4FGL catalog are included for comparison with the distribution above the threshold. \textit{Right:} Detection significance for dark matter subhalos considering five different dark matter masses from 10~GeV (purple) to 1~TeV (orange), at a fixed annihilation cross section of $\sigv = 10^{-23} \, \mathrm{cm}^3\mathrm{s}^{-1}$.}
		\label{fig:TS-hist}
\end{figure}

To determine if a $\gamma$-ray source is classified as \textit{detected}, we adopt the threshold established by the \emph{Fermi}-LAT collaboration in their source analysis \cite{4FGL-DR3}. This threshold corresponds to a detection significance of $5\sigma$ or a TS value of 25. In figure~\ref{fig:TS-hist}, we present the distribution of simulated $\gamma$-ray subhalos as a function of detection significance ($\sigma_d = \sqrt{{\rm TS}}$) for our benchmark setup, in comparison to the distributions of both detected and unidentified (UNID) sources in the 4FGL-DR3 catalog. UNID sources lack clear identification or association with sources at other wavelengths.

For our benchmark dark matter model, 40\% of the original subhalo population is detectable by \emph{Fermi}-LAT. Note that the number of detectable subhalos is significantly enhanced by the large annihilation cross section we assume for the benchmark setup. The $\sigma_d$-distribution for our dark matter subhalos is similar to that of the UNID sources, with a noticeable difference at higher significances ($\sigma_d>10$).

Varying the dark matter mass has a strong effect on the number and distribution of detectable subhalos for a fixed annihilation cross section. Lower masses have a higher detection rate due to a larger overall normalisation and a peak in the annihilation spectrum at lower energies. This effect is quantified in the right panel of Figure~\ref{fig:TS-hist} for our benchmark annihilation cross section $\sigv = 10^{-23}\, \mathrm{cm}^3\mathrm{s}^{-1}$. Annihilation channels other than $b\bar{b}$ or a mixture of channels would further modify the $\sigma_d$ distribution. For example, annihilation into a leptonic final state such as $\tau^+ \tau^-$ would shift the histograms to lower $\sigma_d$ values for each dark matter mass, because the flux peaks at higher photon fluxes.

Our results show that the distribution of the detection significance $\sigma_d$ for a realistic simulation of dark matter subhalos is highly model-dependent, and that it differs from that of 4FGL sources, considering both classified and UNID sources. It roughly approaches that of the 4FGL UNID sources (figure~\ref{fig:TS-hist}, left panel) for specific dark matter masses, a specific annihilation channel and specific cross section values. We thus conclude that enforcing  a $\sigma_d$-distribution that follows that of the UNID sources in the 4FGL catalog regardless of the dark matter model, as done in \cite{Gammaldi:2022wwz}, is a strong simplification that could bias the results of the machine learning classification.   

\subsection{Number of detectable subhalos}

 \begin{figure}[t]
	\center
	\includegraphics[width=.75\textwidth]{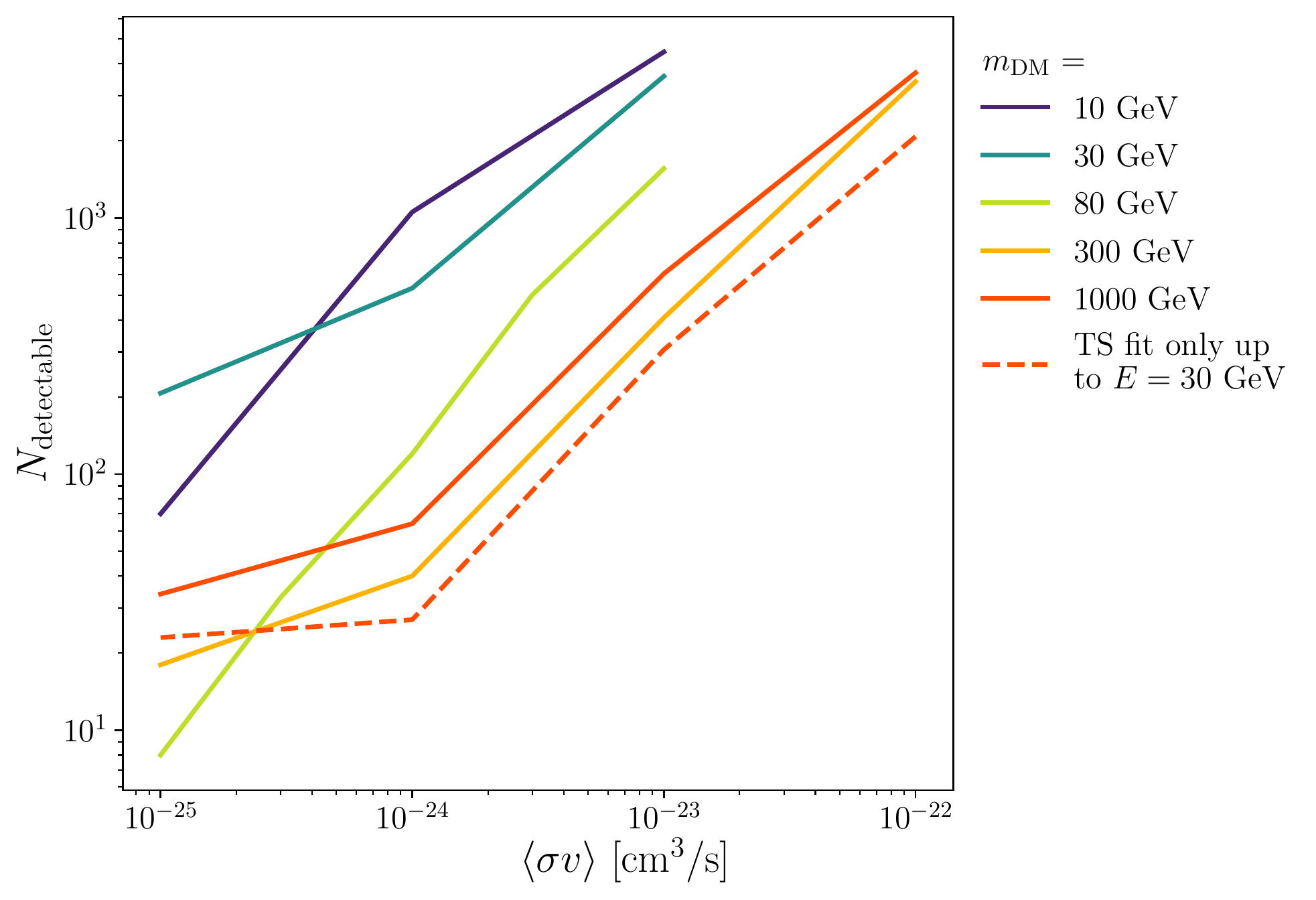}
		\caption{Number of detectable (${\rm TS} \geq 25$) dark matter subhalos in our simulations as a function of the annihilation cross section. The colours represent the  dark matter masses of the respective models. The dashed line for 1~TeV has been obtained by excluding the high energy bins from the sensitivity computation, see text for details.}
		\label{fig:Ndet_sigv}
\end{figure}

The number of detectable subhalos in our simulation setup decreases approximately linearly with decreasing annihilation cross section. This effect has been quantified in previous work~\cite{Bertoni:2015mla,Schoonenberg:2016aml,PhysRevD.96.063009} and is reproduced also in our setup for the 4FGL-DR3 catalog, as illustrated in figure~\ref{fig:Ndet_sigv}. The number of detectable subhalos, i.e.\ subhalos for which  $\sigma_d>5$, ranges from 8 at $\sigv=10^{-25}$ cm$^3$/s to about $10^3$ at $\sigv=10^{-23}$ cm$^3$/s for the benchmark setup.

Figure~\ref{fig:TS-hist} shows that fewer subhalos are detected at higher dark matter masses because the $\gamma$-ray flux decreases proportional to ${\mDM}^{-2}$. However, the decrease in flux is partially offset by the fact that the energy spectrum is harder and thus easier to detect at larger dark matter masses. This is why the number of detectable sources for dark matter masses around 1~TeV is higher than that for 300 GeV, for a given annihilation cross section, as shown in figure~\ref{fig:Ndet_sigv}. This effect has also been observed in a study comparing the 3FGL and 2FHL catalogs \cite{PhysRevD.96.063009}. If we look at the TS obtained by fitting the energy spectrum only up to 30 GeV (red dashed line), we indeed find fewer detectable sources at 1\,TeV compared to dark matter masses of 300\,GeV.

It should be noted that the values in figure~\ref{fig:Ndet_sigv} correspond to a specific realisation of the subhalo distribution and \textsc{fermipy}\ fit, and a statistical uncertainty band should ideally be assigned to each line. However, estimating such a band for the full population of dark matter subhalos, rather than test sources at specific Galactic latitudes, would be computationally prohibitive.

Our analysis shows about two to three times more detectable subhalos than previous work \cite{Bertoni:2015mla,Schoonenberg:2016aml,PhysRevD.96.063009,Coronado-Blazquez:2019pny,Coronado-Blazquez:2019puc}, depending on the dark matter mass. We have identified several differences between our approach and previous analyses that likely contribute to this increased sensitivity. To further validate our results, we performed several tests, which we summarise below.

First, an increase in sensitivity compared to previous analyses is expected because our work is based on the simulation of 12 years of \emph{Fermi}-LAT data, while previous studies have considered the statistics of the 1FGL, 3FGL and 4FGL catalogs (1 year, 4 years and 8 years of data, respectively).

Furthermore, the number of detectable subhalos is calculated using a different strategy to previous analyses. In previous work, such as refs.~\cite{PhysRevD.96.063009,Coronado-Blazquez:2021amj}, a small number of test subhalos are simulated at different sky positions. For each dark matter model (annihilation channel, mass) and latitude bin, the minimum flux (integrated in the considered energy range) for the detection (i.e.\ ${\rm TS}=25$) of such test subhalos is computed.  After compiling this sensitivity, the number of detectable subhalos is obtained by comparing the threshold flux with the expected flux for each subhalo in the population model. In our work, we simulate with \textsc{fermipy}~ each dark matter subhalo obtained for the \texttt{DMonly} population along with the backgrounds, and perform a binned likelihood analysis in each ROI searching for the subhalo emission at its exact simulated position. By fitting each ROI separately, we have more freedom in the ROI model, in particular in the background normalisation and shape, and in the catalog point source parameters. In addition, we exploit the full sensitivity of spectral data analysis in an extended energy range. We recall that this strategy is guided by our goal of simulating the full energy spectrum of each dark matter subhalo in the population, which we use for machine learning classification. 

We have verified a posteriori that the minimum integrated flux we detect is compatible with the threshold computed in previous work. Specifically, for a dark matter mass of $80$\,GeV and  $\sigv = 10^{-25} \, \mathrm{cm}^3 \mathrm{s}^{-1} \,  $ we detect subhalos with ${\rm TS} >25$ and integrated fluxes of $\log_{10}(\phi) \geq -9.15$, which is comparable to the flux sensitivity threshold of ref.~\cite{PhysRevD.96.063009} at $\sim 20^\circ$ with $\log_{10}(\phi_\mathrm{min}) \approx -9.15$. The analysis of ref.~\cite{Coronado-Blazquez:2021amj} presents a slightly higher threshold for point-like sources with $\log_{10}(\phi_\mathrm{min}) \approx -9.69$. This compatibility also holds for larger annihilation cross sections. 

In addition, we have tested that we correctly reconstruct 4FGL-DR3 point sources following the power law with super-exponential cutoff spectral shape. We randomly selected sources in the 4FGL-DR3 with this spectral model and very low ($\sigma_d=5.4$) and high ($\sigma_d=42.1$) significance and latitude ($b=2.7,-74.5$). We then simulated \emph{Fermi}-LAT data using the same configuration as detailed in table~\ref{tab:fermipy_stats} in appendix~\ref{app:fermipy_details}. 
The detection significance and the flux reconstructed by our analysis chain agree well within uncertainties with the values listed in the 4FGL-DR3 catalog in the full energy range. 

\section{Dark matter subhalo classification with neural networks}\label{sec:ml_class}

Our focus now shifts to the unclassified 4FGL-DR3 sources, called UNID, and how they compare with both known astrophysical $\gamma$-ray sources and our sets of detectable dark matter subhalos. Our goal is to identify UNID sources that have the same spectral properties as our dark matter models predict for subhalos, but differ from known astrophysical sources. To achieve this, we use a supervised machine learning classification approach.

\subsection{Algorithm}
We have implemented a Bayesian Neural Network (BNN) \cite{MacKay1992, neal2012bayesian, Gal2016Uncertainty} based on the setup in \cite{Butter:2021mwl} where the advantages of using such a BNN on the energy spectra of $\gamma$-ray sources were shown for classifying blazars of uncertain type. Most notable is the benefit of the uncertainty in the network prediction that can be inferred from the network output. While standard deep neural networks consist of non-linear transformations of discrete values, Bayesian layers allow transformations of probability distributions, as each of the trainable network parameters is assigned an initial weight distribution that is adjusted during training to optimise the network output on the training set. The width of the output distribution is then used to infer the network uncertainty for each sample. The classification is a two-class approach, which we set up in the same way for each individual dark matter model. One class represents the 'dark matter subhalo-like spectra' and the other the already classified 4FGL sources.

\begin{figure}[t]
	\center
	\includegraphics[width=.65\textwidth]{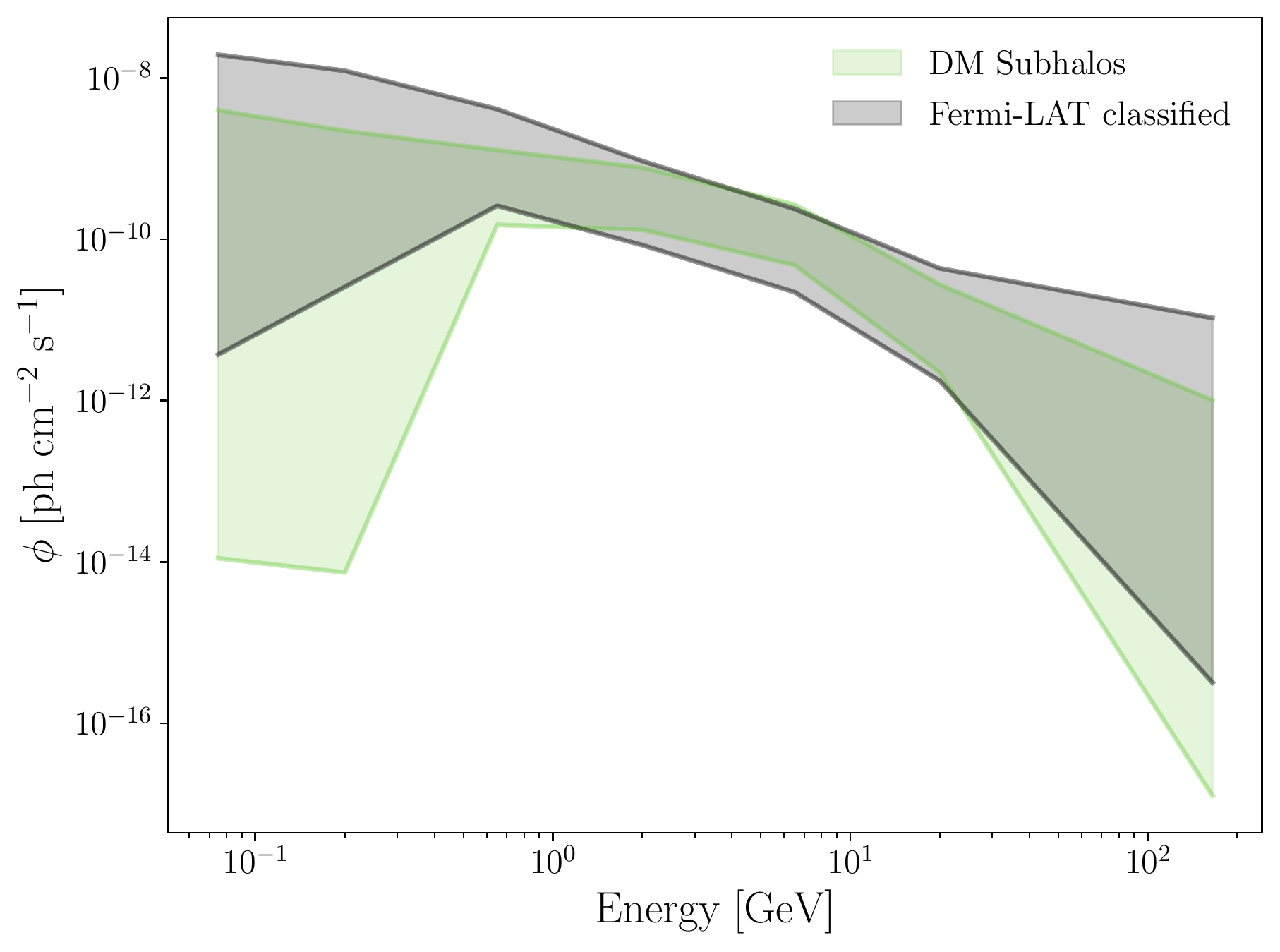}
		\caption{Set of simulated and detectable dark matter subhalo spectra (green) compared to all \emph{Fermi}-LAT classified sources (grey) for our benchmark setup. The shaded regions show the central 68th percentile of the distribution.}
		\label{fig:spectra_plot}
\end{figure}

The training data consists of the 8 values of the source spectra as a function of energy of the 4FGL-DR3 classified sources (\texttt{`Flux\_Band'} in the catalog) and the simulated dark matter subhalos obtained as described in section~\ref{sec:data_sim}. In figure~\ref{fig:spectra_plot} we show the final set of simulated spectra for dark matter subhalos with ${\rm TS}>25$ for our benchmark setup compared to the \emph{Fermi}-LAT classified sources. The shaded region represents the 68th percentile of each distribution. In this particular case, the typical dark matter subhalo spectrum can be very similar to the \emph{Fermi}-LAT classified astrophysical sources, with small deviations at low and high energies. The classifier is designed to be sensitive to these intrinsic features and to take advantage of the full \emph{Fermi}-LAT measurement of the spectrum.

The network consists of four dense Bayesian layers \cite{flipout} with 16 nodes, each with a Gaussian prior on the kernel, ReLU activation and a dropout fraction of 5\,\%. The final layer is fed through a softmax activation. This architecture has proven effective in our setup. For more details on the architecture and optimisation choices see appendix~\ref{app:architecture}.
We use the logarithm of the fluxes as inputs and pre-process them so that for each energy bin the inputs are distributed around a mean of zero with a standard deviation of one. 
For the training process we use the Adam optimiser \cite{adam} with a learning rate of $10^{-3}$ and stop the training when the validation loss converges.

Since the annihilation cross section is only a normalisation factor in our dark matter models, we can assume that the distribution of subhalo spectra is solely determined by the dark matter mass and the annihilation channel. The number of detectable subhalos is mainly dependent on the annihilation cross section, as highlighted in section~\ref{sec:results_det}. Therefore, we only need to create one classifier for each $\mDM$ and annihilation channel. For the purpose of this study, we concentrate on the $\bb$ annihilation channel.

In principle, the number of detectable subhalos is not relevant to the classification approach, as only the shape and normalisation of the spectra are used as network input. However, neural networks can lose accuracy if there is a strong class imbalance in the training set. We therefore set the annihilation cross section for our benchmark model at $\mDM = 80$ GeV and for lower masses to $\sigv = 10^{-23} \, \mathrm{cm}^3 \, \mathrm{s}^{-1}$, and increase this to $\sigv = 10^{-22} \, \mathrm{cm}^3 \, \mathrm{s}^{-1}$ for higher masses to get more statistics in the training set. We note that both of these values are larger than the currently allowed parameter space for WIMP dark matter \cite{Workman:2022ynf}. However, as mentioned above, we only need to ensure a sufficiently large training set, while the network predictions will ultimately depend on the remaining parameters of the model, such as the dark matter mass, the annihilation channel, and the choice of subhalo model. 

The trained BNN provides an estimate of the probability that a source is either of astrophysical origin or a dark matter subhalo corresponding to a particular dark matter model. This classification score ranges from zero to one, with zero corresponding to an astrophysical (4FGL) source and one corresponding to a dark matter subhalo, according to the labels we assign during training.  Since in our BNN approach the network output follows a probability distribution, we can sample an arbitrary number of predictions (generally $10^3$ in our setup) to obtain a mean prediction $\mu$ and the corresponding standard deviation $\sigma$.

We leave 10\% of the training set, not used during training, as test sets. Depending on the similarity of the shape of the subhalo spectra to subsets of the 4FGL sources, the accuracy of our network on the respective test sets ranges from $\sim 90\%$ to $\sim 97\%$ and is consistent for each model with each new training. For the final results, we use the full dataset for training. The lower end of this accuracy range corresponds to the setups with $\mDM = 30$ GeV and $\mDM = 80$ GeV, and is comparable to the accuracy achieved when learning classification between different classes within the 4FGL catalog, as in \cite{Butter:2021mwl}. This further supports our approach and shows that the bias from comparing synthetic data with real data is negligible given the realistic simulation approach we have set up. 

\begin{figure}[t]
	\center
	\includegraphics[width=\textwidth] {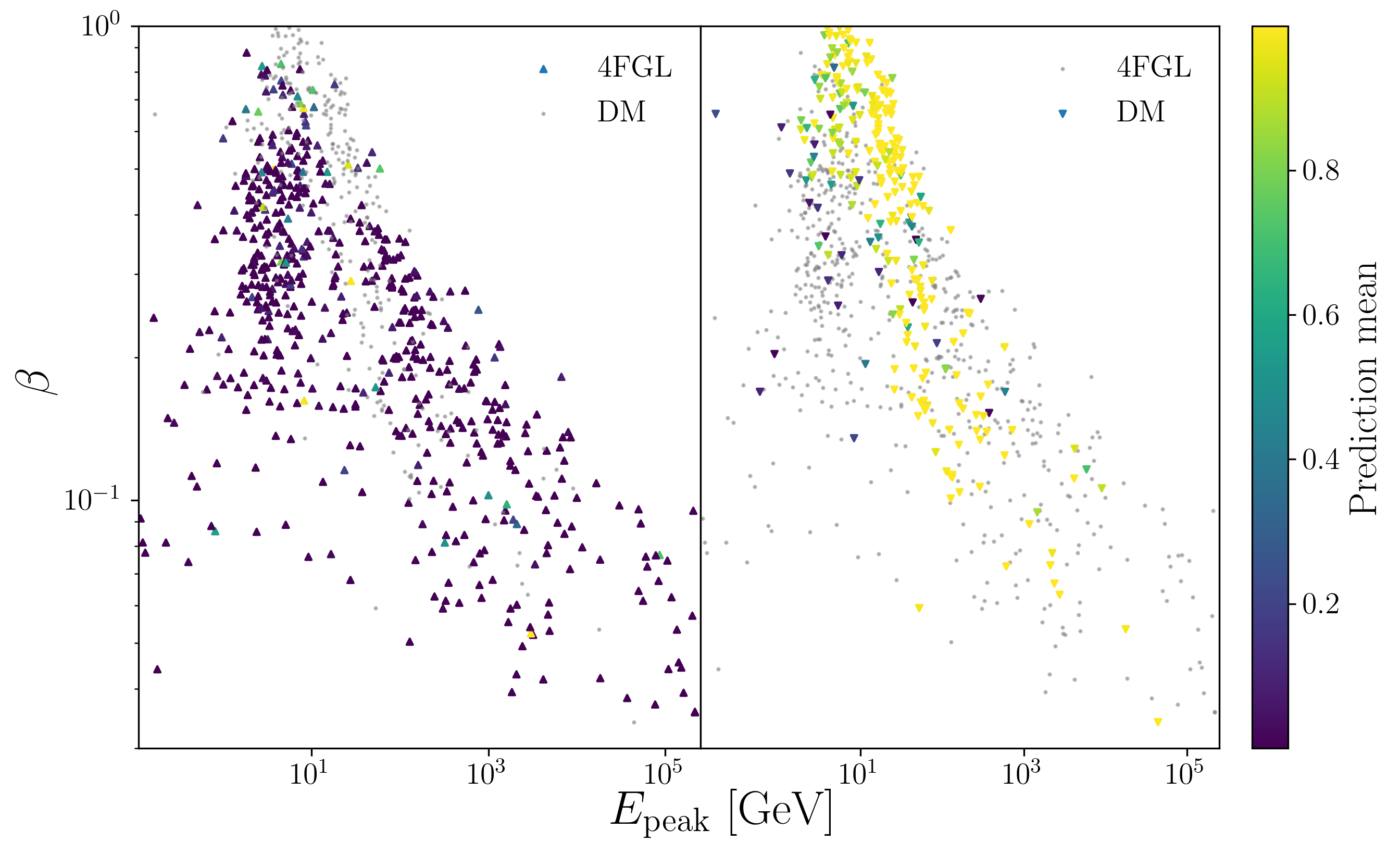} \\
	\includegraphics[width=0.6\textwidth]{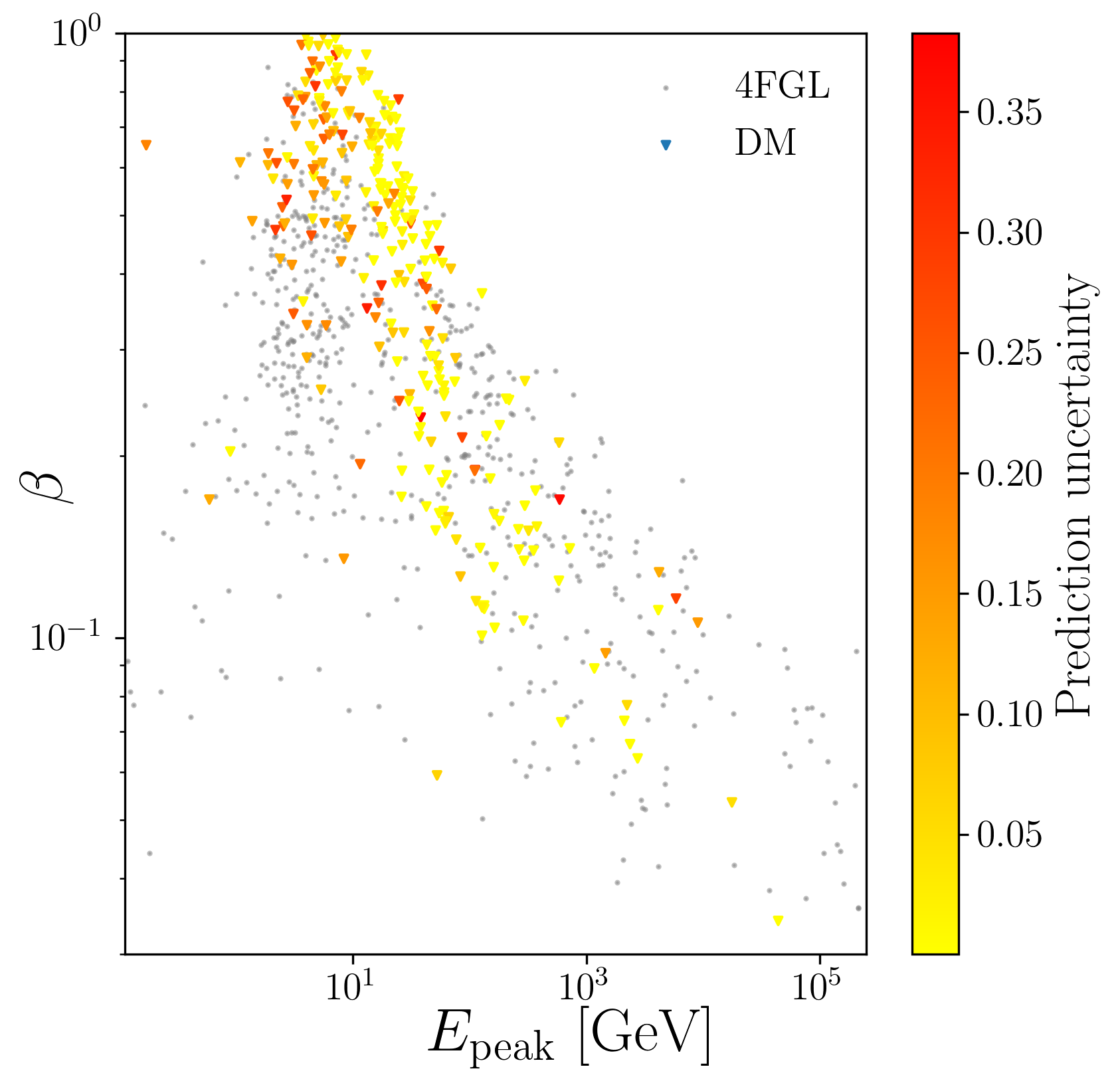}
		\caption{The $\beta$ plot \cite{Coronado-Blazquez:2019pny,Gammaldi:2022wwz}, which is the feature plane of the log-parabola fit parameters $\beta - E_\mathrm{peak}$ of all sources in our test set, is used to illustrate the results of the network for the benchmark setup.\\ \textit{Upper panel:} 4FGL sources (true labels) and dark matter subhalos in the test set. We assign each marker a colour on a scale from 0 to 1, according to the label predicted by the network. The left (right) panel underlines the 4FGL sources (dark matter), while the dark matter subhalos (4FGL) are shown in grey to facilitate comparison. \textit{Lower:} Same as the upper right panel, with the colour scale now showing the uncertainty of the Bayesian network prediction. }
		\label{fig:pred_beta_plot}
\end{figure}

The upper panels in figure~\ref{fig:pred_beta_plot} display the network outputs for the test set of our benchmark model. This figure illustrates how the network accuracy changes with the spectral shape. Since our dataset consists of eight spectrum bins, we simplify the comparison and visualisation of results by using two commonly used \cite{Coronado-Blazquez:2019pny,Gammaldi:2022wwz} parameters -- $\beta$ and $E_{\rm peak}$ -- obtained by fitting a log-parabola: 

\begin{align}
    \phi &= \phi_0 \left( \frac{E}{E_0} \right)^{- \alpha - \beta \log{\left( E / E_0 \right)}} \label{eq:log_parabola} \\
    E_\mathrm{peak} &= E_0\, e^{\frac{2-\alpha}{2 \beta}} \,.
\end{align}

The two upper panels of figure~\ref{fig:pred_beta_plot} display the position of each source in the feature plane, represented by coloured triangular markers. Astrophysical sources are on the left panel, and dark matter subhalos are on the right panel. The colour of the markers corresponds to the mean of the predicted network probabilities. We anticipate that the majority of astrophysical sources will have probabilities near zero, while dark matter subhalos will be near one. Additionally, grey dots indicate the other class, which enables comparison of the two distributions in the feature plane.

This feature plane is commonly used to visualise the different distributions of $\gamma$-ray spectra, and has been exploited for machine learning classification in \cite{Gammaldi:2022wwz}. In our case, we can observe that the two classes learned by the network have distinct features, but there is some overlap in a significant part of the distribution. Misclassification, where the predicted label aligns more with the incorrect class label, is more frequent in these overlapping regions. For example, misclassification is more common around $E_{\rm peak}=10$~GeV, especially for subhalos classified by the network as astrophysical sources. This demonstrates the predictive power of the network for typical spectra and enables us to more cautiously discard outliers.

To further evaluate the network's performance, we analyse the uncertainty of the network's predictions in the feature plane. The results for our benchmark model are presented in the lower panel of figure~\ref{fig:pred_beta_plot}. Here, we only show the samples that correspond to the true label 'dark matter subhalo', and the colour represents the prediction uncertainty. As expected for a Bayesian neural network, we observe that the uncertainty increases for samples that are further away from the denser part of the distribution below $E_{\rm peak}=10$~GeV. Moreover, in this region, we observe larger uncertainties on the predictions for many of the samples that are misclassified, which is also expected. Therefore, we can use the uncertainty to prevent misclassified samples when using the network to make predictions for new spectra. This method is applied to the classification threshold that is introduced in the following section.

\subsection{Results}

We use the classification algorithm introduced in the previous section to estimate the number of $\gamma$-ray spectra among the UNID sources observed by \emph{Fermi}-LAT that exhibit characteristics similar to those expected from dark matter subhalos. 

We restrict our analysis to UNID sources, i.e.\ detected sources that have not been classified as any of the astrophysical source types, and apply conservative selection criteria to avoid counting sources that can be reasonably discarded as subhalo objects. We start with 2157 UNID sources from the 4FGL-DR3 catalog and remove all sources with a variability index greater than 24.725, corresponding to a 99\% confidence level that the source is variable in time \cite{4FGL-DR3}. This cut leaves us with 2067 candidates. We also discard sources with a 'low confidence association' with any astrophysical source type, leaving 1788 UNID sources for classification with our BNN. 

We further distinguish on-plane and off-plane sources based on their location within and outside the Galactic plane, respectively, using the threshold $|b| = 10^{\circ}$. This yields 890 on-plane and 898 off-plane sources for analysis, allowing us to compare the on-plane and off-plane fractions within the classification of the UNID sources.
We expect the dark matter subhalos to be more evenly distributed across the Galactic halo, rather than clustering predominantly in the Galactic plane. In particular, because we expect pulsars with an energy spectrum similar to that expected from dark matter annihilation to be predominantly in the Galactic plane, the number of off-plane sources classified as subhalos is a more conservative estimate.

We use the class predictions from our network to distinguish between astrophysical sources and dark matter subhalos. By setting a threshold on the mean prediction $\mu$ minus the standard deviation $\sigma$, we can estimate the number of potential dark matter subhalos among the UNID sources. We apply a range of thresholds ($\mu - \sigma \geq [0.5, 0.6, 0.7, 0.8]$) to capture the behaviour of the predictions within a loose and tight selection threshold, as described in \cite{Butter:2021mwl}. Comparing the different thresholds allows us to understand the range of candidates, from a more conservative lower limit to a more optimistic upper limit.

\begin{figure}[t]
    \centering
    \includegraphics[width=0.7\textwidth]{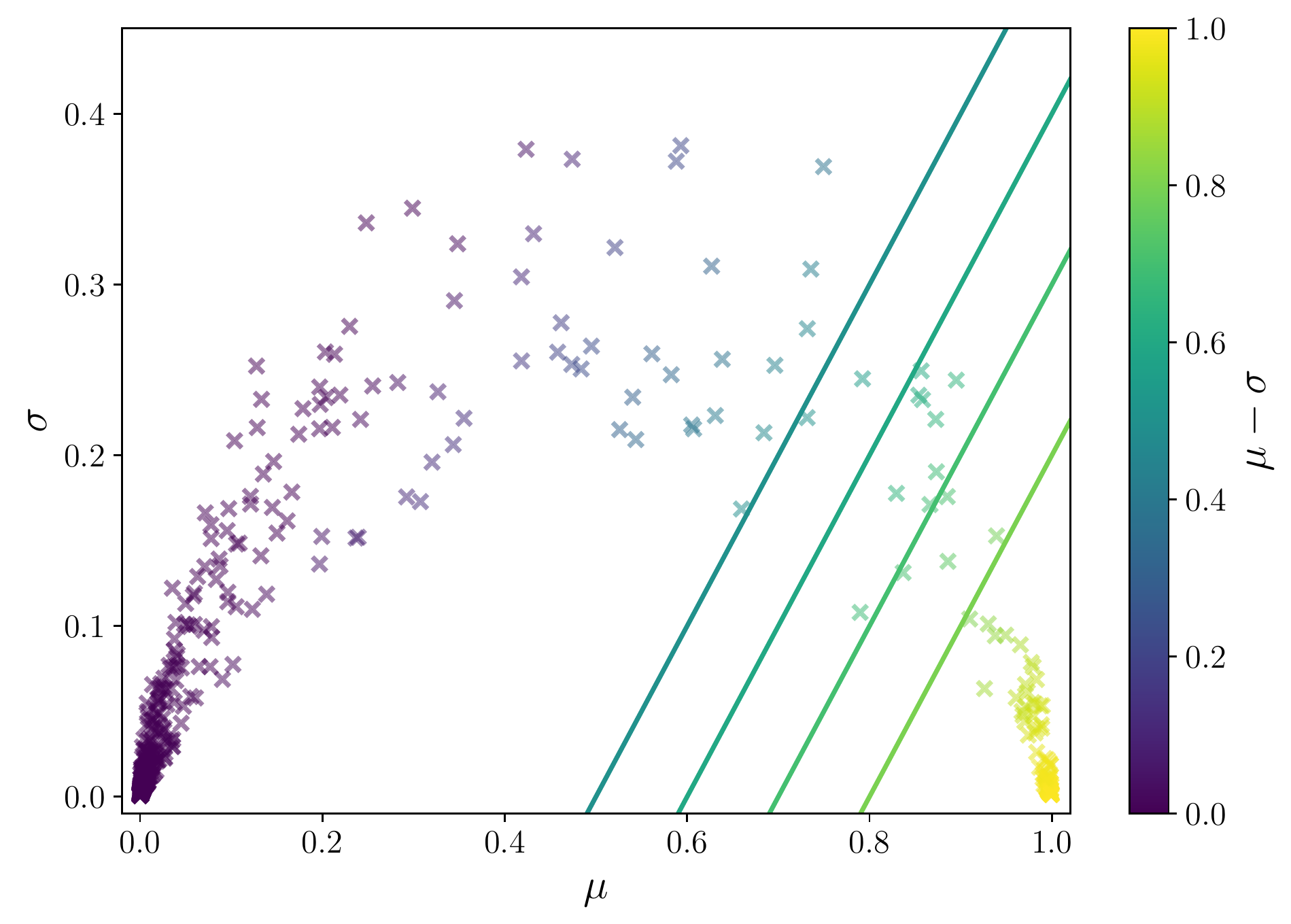}
    \captionof{figure}{Distribution of mean $\mu$ and uncertainty $\sigma$ for the test set of the benchmark setup. The color gradient shows the $\mu - \sigma$ quantity that we use to define the classification threshold. Sources located in the lower right corner of the plane are classified as dark matter subhalos with high confidence, while likely astrophysical sources are located in the lower left. The solid lines mark the different choices for the threshold used to select the candidates.}  
    \label{fig:rainbow_threshold}
\end{figure}

We illustrate this approach in figure~\ref{fig:rainbow_threshold} for the test set of our benchmark setup. The figure shows the mean prediction $\mu$ and standard deviation $\sigma$ for each source, with the colour gradient indicating the difference $\mu - \sigma$. The thresholds are shown as straight lines $\mu - \sigma$ in the figure. Most sources in this test set are astrophysical, so they are consistently classified as such with a mean prediction close to zero and a low uncertainty $\sigma<0.1$. Sources that are not easily distinguishable by the network have a high uncertainty, as expected. To incorporate this uncertainty, we use it in our classification threshold. By increasing the threshold, we increase the fraction of sources classified with high confidence as dark matter subhalos, resulting in a smaller overall number of candidates.

We use a threshold to determine the number of subhalo candidates, i.e.\ UNID sources that survive our cuts and have a predicted class label above the threshold. These candidates are particularly relevant to the corresponding dark matter model, as their spectra are consistent with the subhalo model with a certain level of confidence set by the threshold. While the network predictions do not confirm the detection of new physics, they do indicate the dark matter model with which the spectra are most consistent. However, there may be a degree of degeneracy, as shown in the feature plane in figure~\ref{fig:pred_beta_plot}. Nevertheless, this list of subhalo candidates can serve as a first indication of UNID sources that may be promising for further investigation, especially in a multi-wavelength approach.

\begin{figure}[t]
    	\centering
    	\includegraphics[width=0.44\textwidth]{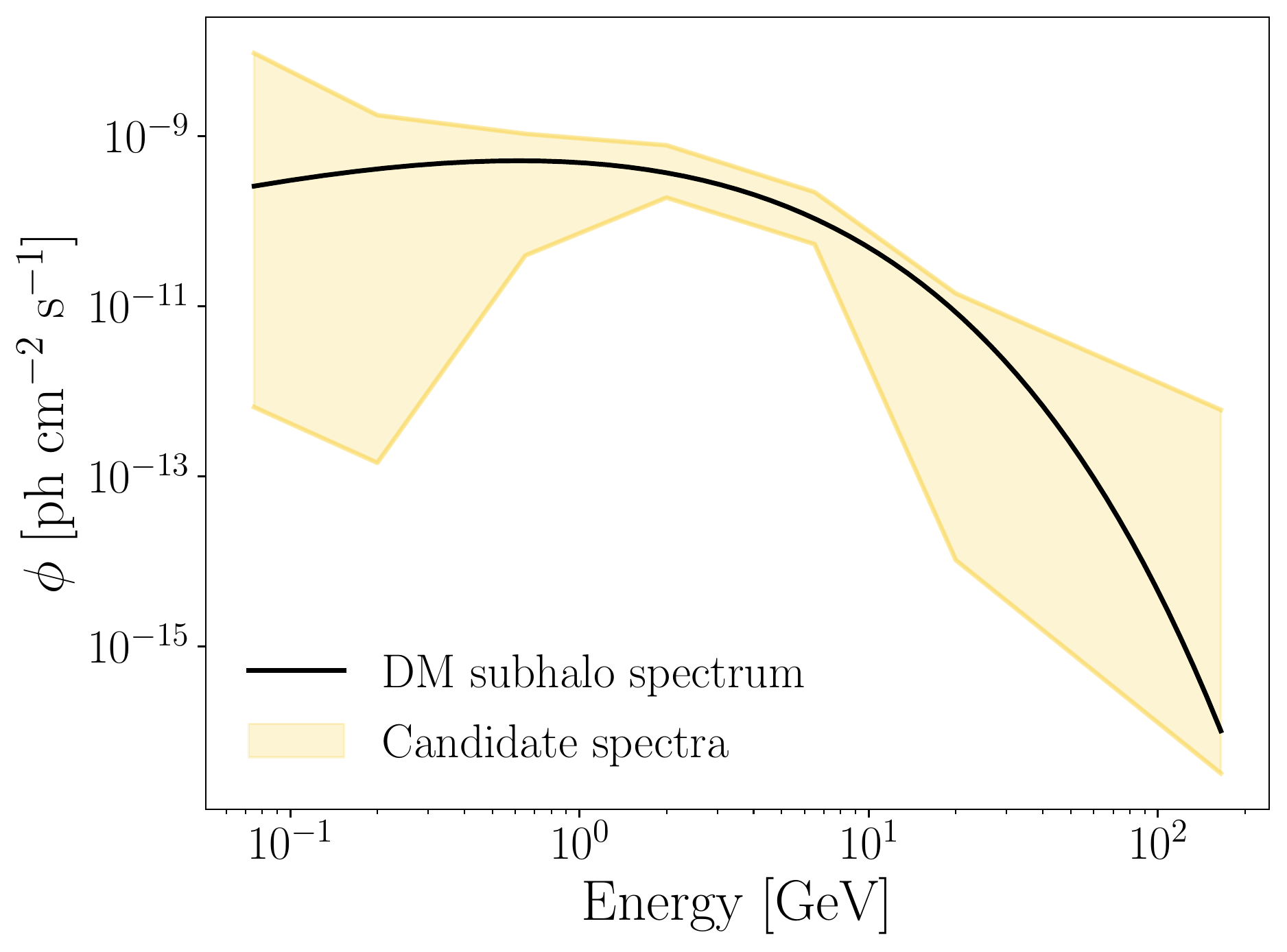}
        \includegraphics[width=0.54\textwidth]{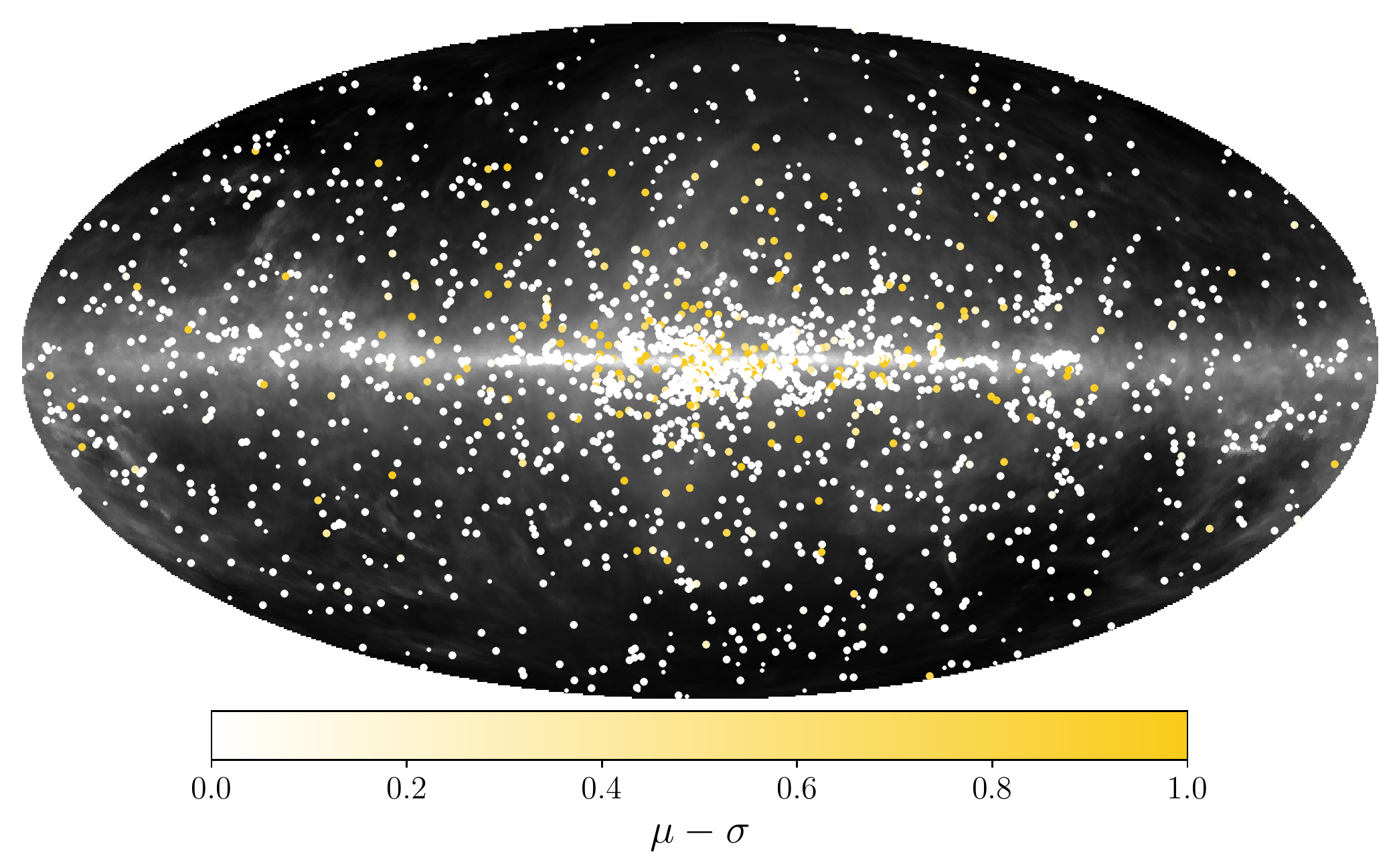}
    		\caption{\textit{Left:} 68th percentile of the distribution of the spectra of the UNID sources classified by our networks as subhalo candidates for the benchmark model ($m_{\rm}=80$~GeV, $\mu-\sigma=0.5$) compared to the theoretical spectrum. \textit{Right:} Skymap showing the position of the UNID sources in the 4FGL-DR3. The colour of the marker indicates the $\mu-\sigma$ as obtained by our neural network, where $\mu-\sigma\sim1$ are sources with spectra more similar to dark matter for the benchmark model.}
    		\label{fig:candidate plots}
\end{figure}

Table~\ref{tab:Ncandidate} shows the number of candidates individually for on-plane and off-plane sources for different dark matter masses and thresholds. If we assume that there is a non-negligible population of dark matter subhalos already detected among the \emph{Fermi}-LAT $\gamma$-ray sources, their spectrum could match well with that expected from the annihilation of dark matter particles with masses of a few tens of GeV. This is illustrated in the left panel of figure~\ref{fig:candidate plots}, where we show the 68th percentile of the distribution of the candidate spectra for the benchmark setup and a threshold of $\mu-\sigma=0.5$, compared to the theoretical prediction for the dark matter subhalo spectrum (eq.~\eqref{eq:flux}). 

The skymap in the right panel of the figure~\ref{fig:candidate plots} shows the position of all UNID sources in the 4FGL-DR3. The sources that are not considered in our classification (e.g. due to time variability or low probability of association) are indicated by smaller markers. The colour of the marker indicates the value of the network prediction for the quantity $\mu-\sigma$. High confidence subhalo candidates (dark yellow) are approximately isotropic distributed across the sky and are not concentrated in regions of high Galactic background, shown in grey as in figure~\ref{fig:Jfac_skymap}. We make the network predictions for the UNID sources publicly available for each model.\footnote{\url{https://github.com/kathrinnp/bnn-subhalo-candidates}}

\begin{table}[t]
\setlength{\tabcolsep}{1pt}
 \caption{Number of dark matter subhalo candidates in the 4FGL UNID at different classification thresholds. The number of off-plane sources is shown in black, and on-plane sources are shown in grey.}
    \centering
    \begin{tabular}{ccrclcrclcrclcrcl}
    $\,$ & $\,$ \\
    \toprule
      $\mDM$ & $\qquad$ & \multicolumn{15}{c}{$N_\mathrm{candidate}$ at threshold}  \\
      &  $\qquad$ & \multicolumn{3}{c}{0.5} & & \multicolumn{3}{c}{0.6} & & \multicolumn{3}{c}{0.7} & &\multicolumn{3}{c}{0.8} \\
         \midrule
        10 GeV & $\qquad$
        & 8 & {\color{gray} +} 
        & {\color{gray} 37} & $\quad$
        & 6 
        & {\color{gray} +} 
        & {\color{gray} 32} & $\quad$
        & 4 
        & {\color{gray} +} 
        & {\color{gray} 28} & $\quad$
        & 4 
        & {\color{gray} +} 
        & {\color{gray} 23}
        \\
        30 GeV & $\qquad$
        & 121 
        & {\color{gray} +} 
        & {\color{gray} 181} & $\quad$
        & 111 
        & {\color{gray} +} 
        & {\color{gray} 164} & $\quad$
        & 97 
        & {\color{gray} +} 
        & {\color{gray} 146} & $\quad$
        & 88 
        & {\color{gray} +} 
        & {\color{gray} 127}
        \\
        80 GeV & $\qquad$
        & 77 
        & {\color{gray} +} 
        & {\color{gray} 111} & $\quad$
        & 67 
        &{\color{gray} +} 
        & {\color{gray} 103} & $\quad$
        & 64 
        & {\color{gray} +} 
        & {\color{gray} 94} & $\quad$
        & 49 
        & {\color{gray} +} 
        & {\color{gray} 85}
        \\
        300 GeV & $\qquad$
        & 22
        & {\color{gray} +} 
        & {\color{gray} 66} & $\quad$
        & 19 
        & {\color{gray} +} 
        & {\color{gray} 56} & $\quad$
        & 11
        &{\color{gray} +} 
        & {\color{gray} 54} & $\quad$
        & 8
        & {\color{gray} +} 
        & {\color{gray} 41}
        \\
        1 TeV & $\qquad$
        & 1
        & {\color{gray} +} 
        & {\color{gray} 10} & $\quad$
        & 0
        & {\color{gray} +} 
        & {\color{gray} 4} & $\quad$
        & 0 
        & {\color{gray} +} 
        & {\color{gray} 2} & $\quad$
        & 0
        & {\color{gray} +} 
        & {\color{gray} 1} 
        \\
        \bottomrule
    \end{tabular}
    \label{tab:Ncandidate}

\end{table}

\section{Limits on the dark matter annihilation cross section }
\label{sec:limits}

We can use our classification algorithm to set upper limits on the dark matter annihilation cross section. To do this, we distinguish between sources that are confidently astrophysical in origin and those that are considered exotic. Exotic sources include $\gamma$-ray sources that are confidently identified as subhalos, as well as sources that cannot be attributed to either dark matter subhalos or astrophysical objects. We select these sources using a threshold $t$ such that $(1-\mu)-\sigma\leq t$. The quantity $1-\mu$ corresponds to the classification score for astrophysical sources, and switching the labels provides a more conservative selection of sources. As the threshold increases, more sources are not classified as astrophysical with sufficient confidence, leading to an increase in the number of exotic sources. The number of exotic sources for our models is listed in table~\ref{tab:Nexotic}.

\begin{table}[t]
    \setlength{\tabcolsep}{1pt}
    \centering
    \caption{Same as table~\ref{tab:Ncandidate} with numbers representing non-astrophysical (\textit{`exotic'}) sources according to the network prediction.}
    \begin{tabular}{ccrclcrclcrclcrcl}
    $\,$ & $\,$ \\
    \toprule
      $\mDM$ & $\qquad$ & \multicolumn{15}{c}{$N_\mathrm{exotic}$ at threshold}  \\
      &  $\qquad$ & \multicolumn{3}{c}{0.5} & & \multicolumn{3}{c}{0.6} & & \multicolumn{3}{c}{0.7} & &\multicolumn{3}{c}{0.8} \\
         \midrule
        10 GeV & $\qquad$
        & 28 & {\color{gray} +} 
        & {\color{gray} 95} & $\quad$
        & 28 
        & {\color{gray} +} 
        & {\color{gray} 112} & $\quad$
        & 36
        & {\color{gray} +} 
        & {\color{gray} 128} & $\quad$
        & 41
        & {\color{gray} +} 
        & {\color{gray} 146}
        \\
        30 GeV & $\qquad$
        & 224 
        & {\color{gray} +} 
        & {\color{gray} 319} & $\quad$
        & 236 
        & {\color{gray} +} 
        & {\color{gray} 347} & $\quad$
        & 254
        & {\color{gray} +} 
        & {\color{gray} 366} & $\quad$
        & 281
        & {\color{gray} +} 
        & {\color{gray} 410}
        \\
        80 GeV & $\qquad$
        & 140 
        & {\color{gray} +} 
        & {\color{gray} 161} & $\quad$
        & 159
        &{\color{gray} +} 
        & {\color{gray} 177} & $\quad$
        & 191
        & {\color{gray} +} 
        & {\color{gray} 191} & $\quad$
        & 233
        & {\color{gray} +} 
        & {\color{gray} 211}
        \\
        300 GeV & $\qquad$
        & 66
        & {\color{gray} +} 
        & {\color{gray} 105} & $\quad$
        & 80 
        & {\color{gray} +} 
        & {\color{gray} 110} & $\quad$
        & 98
        &{\color{gray} +} 
        & {\color{gray} 125} & $\quad$
        & 126
        & {\color{gray} +} 
        & {\color{gray} 137}
        \\
        1 TeV & $\qquad$
        & 16
        & {\color{gray} +} 
        & {\color{gray} 50} & $\quad$
        & 30
        & {\color{gray} +} 
        & {\color{gray} 87} & $\quad$
        & 43 
        & {\color{gray} +} 
        & {\color{gray} 112} & $\quad$
        & 61
        & {\color{gray} +} 
        & {\color{gray} 134} 
        \\
        \bottomrule
    \end{tabular}
    \label{tab:Nexotic}
\end{table}

To place a conservative upper limit on the dark matter annihilation cross section, we adopt the criterion that the number of detectable subhalo candidates should not exceed the number of exotic sources, i.e.\ those sources that cannot be confidently classified as astrophysical. Using the relationship between the annihilation cross section $\sigv$ and the number of detectable subhalo candidates (as shown in figure~\ref{fig:Ndet_sigv}), we can translate the limit on the number of subhalo candidates into an upper limit on the annihilation cross section $\sigv$.

\begin{figure}[t]
	\center
	\includegraphics[width=.65\textwidth]{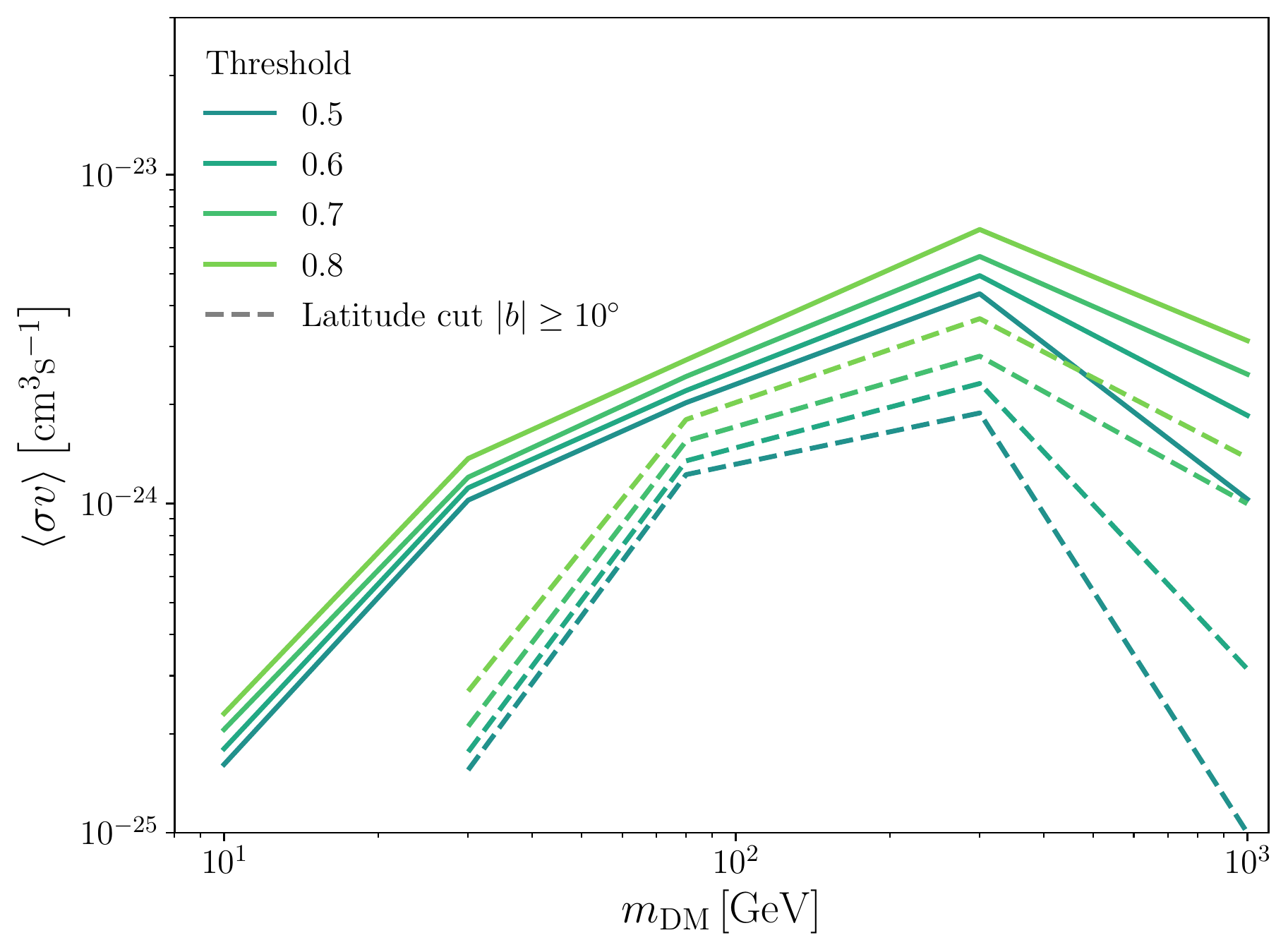}
		\caption{Limits on the annihilation cross section as a function of dark matter mass, assuming annihilation into $\bb$. The limits are obtained by matching the number of detectable subhalos in our simulation with the number of all subhalo candidates \textit{(solid lines)} and candidates at  latitudes $|b| \leq 10^{\circ}$\textit{(dashed lines)} in the 4FGL UNID sources for each classification threshold we consider (colors as in figure~\ref{fig:rainbow_threshold}).}
		\label{fig:sigv_limits_mDM}
\end{figure}

The resulting limits on the dark matter annihilation cross section are shown in figure~\ref{fig:sigv_limits_mDM} for different classification thresholds as a function of the dark matter mass. As expected, more stringent classifications of the spectra as astrophysical lead to weaker limits. 

In figure~\ref{fig:sigv_limits_mDM}, the upper limits on the dark matter annihilation cross section are shown using all exotic sources (solid lines) and only off-plane sources (dashed lines). Since the off-plane sources are expected to have less contamination from astrophysical sources, they are a better representation of the potential dark matter subhalos. However, the latitude cut applied to obtain the off-plane sample also removes a significant number of UNID sources. Therefore, the all-sky results should be considered conservative, while the off-plane results represent a complementary approach with less contamination from pulsars or other astrophysical $\gamma$-ray sources.

Our results show particularly strong limits for the large dark matter mass model $\mDM = 1$\,TeV, with $\sigv< 10^{-24}$ cm$^{3}$s$^{-1}$ excluded for the loose selection. As discussed in section~\ref{sec:results_det}, subhalos with dark matter particles of this mass have $\gamma$-ray spectra that peak at very high energies, making them easier to detect above the background with high significance. Moreover, these spectra are clearly distinguishable from the majority of UNID sources, which follow the same distribution as the classified 4FGL sources. Consequently, there are few exotic sources remaining in the direct comparison performed by the network with the 1 TeV model, resulting in strong and well-motivated limits. As discussed in section~\ref{sec:results_det}, excluding the highest energies in the spectra for the TS fit, where the background statistics are more pronounced, weakens the limits for the 1 TeV model, as shown in figure~\ref{fig:sigv_limits_mDM_1TeV} in appendix~\ref{app:1tev}.

\begin{figure}[t]
	\center
	\includegraphics[width=.65\textwidth]{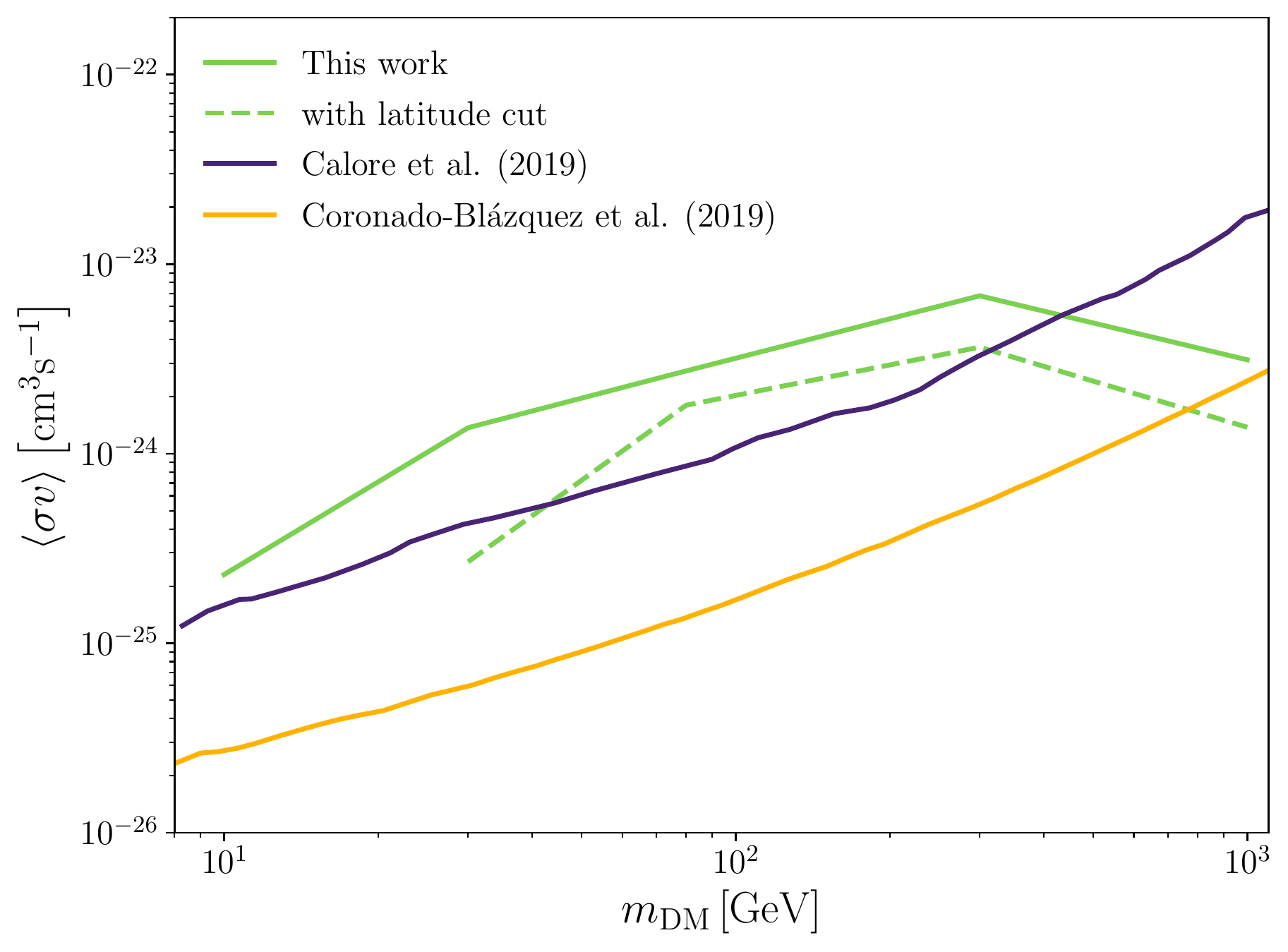}
		\caption{Comparison of our most conservative limits ($t=0.8$), as in figure~\ref{fig:sigv_limits_mDM}, with other work searching for dark matter subhalos in \emph{Fermi}-LAT data \cite{Calore:2019lks,Coronado-Blazquez:2019pny}.}
		\label{fig:sigv_limits_compare}
\end{figure}

Our results can be compared with previous research on dark matter subhalo searches, such as the studies by Coronado-Blazquez et al. (2019a,b) \cite{Coronado-Blazquez:2019pny, Coronado-Blazquez:2019puc} and Calore et al. (2019) \cite{Calore:2019lks}, which we include in figure~\ref{fig:sigv_limits_compare}. We present our most conservative limits, where we assume that the neural network must be highly confident in classifying a source as astrophysical, with $(1-\mu)-\sigma\leq t=0.8$, resulting in a large number of potentially exotic sources. Therefore, the limits we obtain are weaker than those of other studies in the central energy range, but become stronger and more competitive at low and high dark matter masses, where the subhalo spectra differ more from astrophysical sources. Unlike previous studies where the number of candidates was fixed for all dark matter masses studied, we vary the number of sources as a function of dark matter mass, highlighting the advantage of using $\gamma$-ray energy spectra and the importance of a well-trained neural network on known astrophysical sources, as well as carefully constructed subhalo simulations.
In addition, we anticipate more constraining results for annihilation channels for which the dark matter spectrum has a more distinct shape with respect to observed astrophysical source, e.g. $\tau^+ \tau^-$. 
Finally, we stress again that our criterion based on $N_{\rm exotic}$ is a conservative choice. When considering only the sources included in the $N_{\rm candidate}$ set (compare table~\ref{tab:Ncandidate} with table~\ref{tab:Nexotic}), stronger constraints (up to a factor of five) can be derived.

\section{Conclusions}\label{sec:conclusions}

The search for dark matter subhalos is an exciting aspect of the indirect search for dark matter. Subhalos are expected to form within galactic halos, such as that of the Milky Way, through hierarchical clustering of substructures. Subhalos can produce radiation, such as $\gamma$ rays, through the annihilation of dark matter particles. We focus on $\gamma$ rays with energies in the GeV-TeV range, which can be directly traced to their source if detected on Earth.

The \emph{Fermi}-LAT collaboration has provided the most comprehensive catalog of $\gamma$-ray point sources in our Galaxy, with about 6,600 sources in the latest data release, 4FGL-DR3. About a third of these sources cannot be confidently associated with any known astrophysical object and may represent new, exotic sources, such as dark matter subhalos. To search for subhalo-like objects in the unidentified \emph{Fermi}-LAT sources, we have developed a new analysis framework based on supervised neural networks. 

To model the dark matter subhalo population, we used the \texttt{CLUMPY} code based on dark matter-only cosmological simulations. This provided us with the position and $\Jcal$-factor distribution of about 4500 Galactic subhalos. We then set up a range of dark matter spectral models by varying the dark matter mass, assuming that the $\gamma$-ray flux is produced by dark matter annihilation into $\bb$ final states. 

We calculated the expected flux in the \emph{Fermi}-LAT energy range for each subhalo using the $\gamma$-ray energy spectrum from dark matter annihilation and the $\Jcal$ factor of each subhalo. To compare the modelled subhalos with the observed $\gamma$-ray spectra, we simulated the measurement of $\gamma$ rays with \emph{Fermi}-LAT using the \textsc{fermipy} tool, which takes into account all relevant backgrounds and detected point sources. This allowed us to obtain realistic subhalo spectra that follow the same statistical and systematic distributions as the catalog, and to exclude subhalos that are not detected with sufficient significance. 

Our first new result is an estimate of the detectability of dark matter subhalos with different dark matter models, which we have updated to the statistics of 12 years of \emph{Fermi}-LAT data. We found that the detectability decreases with increasing dark matter mass due to the spectral form of the dark matter annihilation fluxes. However, at larger masses, the spectra peak at energies where the astrophysical $\gamma$-ray background is less prevalent, increasing the subhalo detectability.

The main innovation of this study is the use of Bayesian neural networks to classify the unidentified \emph{Fermi}-LAT sources. The network was trained using both the astrophysically classified \emph{Fermi}-LAT sources and the set of detectable dark matter subhalo spectra that we obtained for each model. To achieve this, we trained one network configuration per model and carried out careful testing to ensure that the classifier had a high accuracy of at least 90\%, while avoiding any bias that might arise from comparing simulated and real data. Each prediction made by our network provides an indication of similarity to one of the two trained classes, as well as an uncertainty measure, which enables accurate candidate selection.

Our use of Bayesian neural networks to classify unidentified \emph{Fermi}-LAT sources has led to the identification of numerous dark matter subhalo candidates, particularly in the mass range of a few tens of GeV. We are making this list of candidate sources publicly available to allow further investigation of their nature using multi-wavelength observations.

In addition, we have used the number of $\gamma$-ray sources that the network could not confidently classify as astrophysical to derive conservative upper bounds on the dark matter annihilation cross sections. Our limits are particularly competitive at large dark matter masses, where subhalo spectra are more distinct from astrophysical sources. We therefore demonstrate the importance of separately evaluating the number of subhalo candidates in \emph{Fermi}-LAT catalogs for each dark matter model in order to fully exploit the information contained in the spectra.

Further work could explore more sophisticated models of subhalo formation, including for example baryonic effects, or more complex particle physics scenarios involving alternative or multiple annihilation channels. In addition, the application of unsupervised and weakly supervised machine learning techniques could enable the identification of anomalous $\gamma$-ray sources in a more model-independent way. 

Our method of combining simulated $\gamma$-ray spectra with observed data in machine learning classification tasks can be used for similar applications. Since the size of observed $\gamma$-ray datasets is limited, data augmentation techniques are needed to classify small datasets or to analyse new, unobserved sources. Our approach can improve such techniques and make them more effective.

\acknowledgments
We thank Francesca Calore for reading the manuscript and providing insightful comments, and Paolo Salucci for helpful comments on the dark matter profile. 
A.B. would like to acknowledge support by the Deutsche Forschungsgemeinschaft (DFG,
German Research Foundation) under grant 396021762 – TRR 257 Particle Physics Phenomenology after the Higgs Discovery and by the BMBF for the AI junior group 01IS22079.
A.B. gratefully acknowledges the continuous support from LPNHE, CNRS/IN2P3,
Sorbonne Université and Université de Paris.
S.M. acknowledges the European Union's Horizon Europe research and innovation programme for support under the Marie Sklodowska-Curie Action HE MSCA PF–2021, grant agreement No.10106280, project \textit{VerSi}.
Simulations and neural network training were performed with computing resources granted by RWTH Aachen University under project `rwth0754'.

\medskip 

\appendix

%

\section{Technical notes}

\subsection{Fit to dark matter $\gamma$-ray spectra}\label{app:spec}
In this section we report additional details on the fit to the dark matter $\gamma$-ray spectra with the functional form defined in eq.~\eqref{eq:PLSuperExpCutoff}, which can be used within the \textsc{fermipy} framework. PLSuperExpCutoff is a fitting function that allows for control over the parameters (prefactor $N_0$, cutoff energy $E_c$, and indices $\Gamma$ and $\beta$) and the uncertainty of the fit. We compare this fit to the original energy spectra as taken from \cite{Cirelli:2010xx} and the DMFitFunction \cite{Jeltema:2008hf}, a tabulated fit to an earlier computation of the dark matter annihilation spectra also available in \textsc{fermipy}. 

\begin{figure}[t]
	\center
	\includegraphics[width=.65\textwidth]{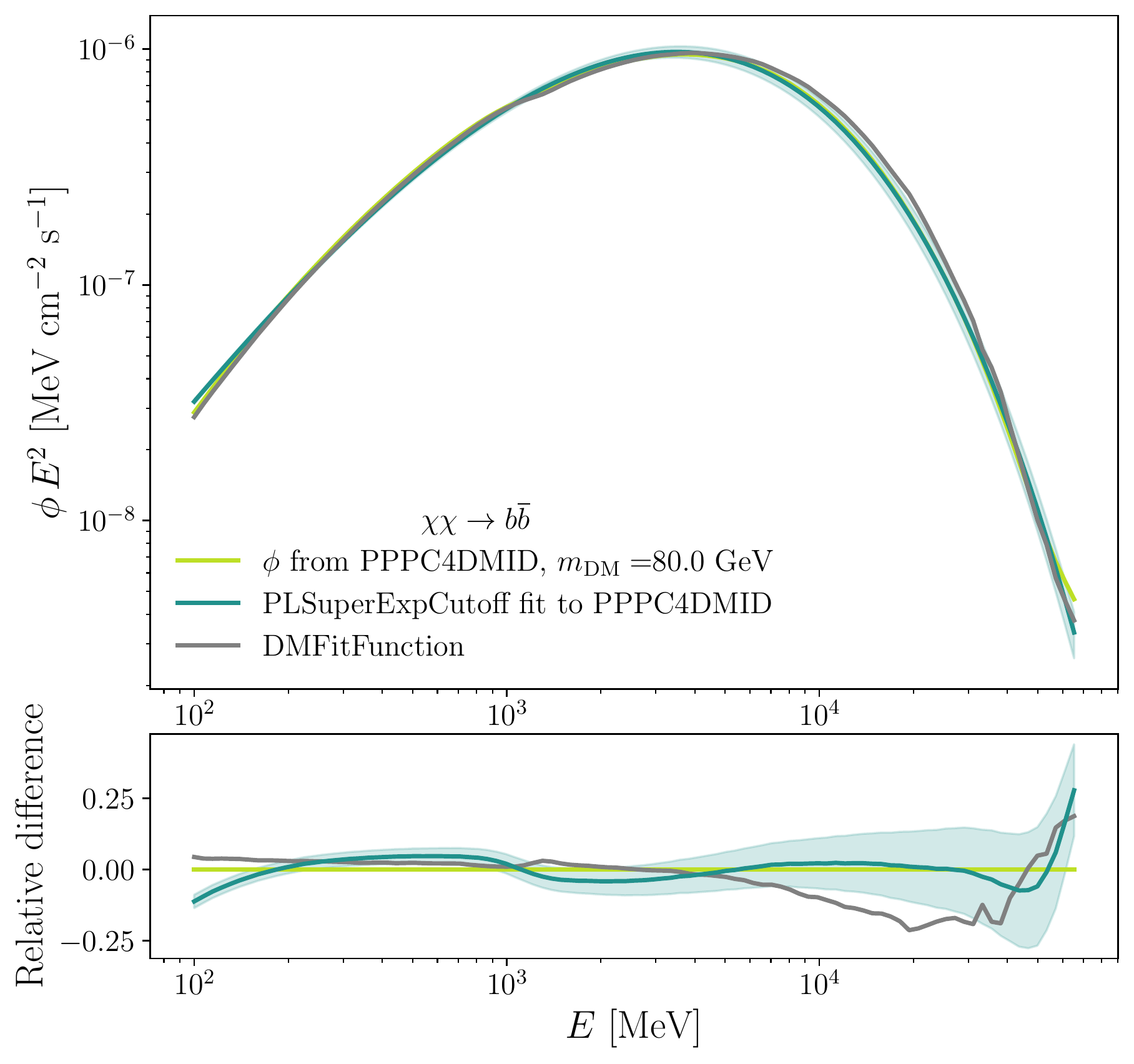}
		\caption{Fit to the dark matter annihilation spectrum in the $b\bar{b}$ channel for a dark matter mass of 80~GeV from \cite{Cirelli:2010xx} (\textit{yellow}). We compare two options for fitting such a spectrum: our fit using the PLSuperExpCutoff function (\textit{green}) and the DMFitFunction available in \textsc{fermipy}\, (\textit{grey}). The lower inset shows the relative difference of the fit functions to the original spectrum. The shaded band for the PLSuperExpCutoff represents the $1 \sigma$ uncertainty band.}
		\label{fig:DMFitFunction}
\end{figure}

In figure~\ref{fig:DMFitFunction} we show the spectra for these three cases for an example dark matter mass of 80~GeV in the default $b\bar{b}$ annihilation channel. Very similar results are obtained for the other dark matter masses studied in this paper. The lower inset in each panel shows the relative difference of the fit with PLSuperExpCutoff and DMFitFunction with respect to the spectra extracted from \cite{Cirelli:2010xx}. 
The PLSuperExpCutoff fit agrees well with the original spectra in the full energy range, with relative differences below a few percent below 10~GeV and reaching 20-25\% at the highest energies. 
We note that using the fit instead of the original spectra does not introduce significant spectral distortions. It is therefore safe to use the PLSuperExpCutoff fit to simulate the energy spectra of dark matter subhalos and for classification with our networks. Similar differences are found for the DMFitFunction. 

\subsection{Simulation of \emph{Fermi}-LAT data}
\label{app:fermipy_details}

We simulate 12 years of data from 4 August 2008 to 2 August 2020 in the energy range between 50~MeV and 1~TeV, an interval compatible with that used to compile the 4FGL-DR3 catalog. We select Pass~8 \texttt{P8R3\_SOURCE} events, selecting all available photons (event type FRONT + BACK) and use consistently the \texttt{P8R3\_SOURCE\_V3} instrument response function.  Standard quality selection criteria are applied, i.e.\ DATA\_QUAL>0 \&\& LAT\_CONFIG==1 and a maximum  zenith angle of $105$ degrees.\footnote{\url{https://fermi.gsfc.nasa.gov/ssc/data/analysis/documentation/Cicerone/Cicerone_Data_Exploration/Data_preparation.html}} Unless otherwise stated, we bin the data using an angular pixel size of $0.5$~deg  and three energy bins per decade in energy, for a total of 13 bins covering the analysis interval.

\begin{figure}[t]
	\center
	\includegraphics[width=.95\textwidth]{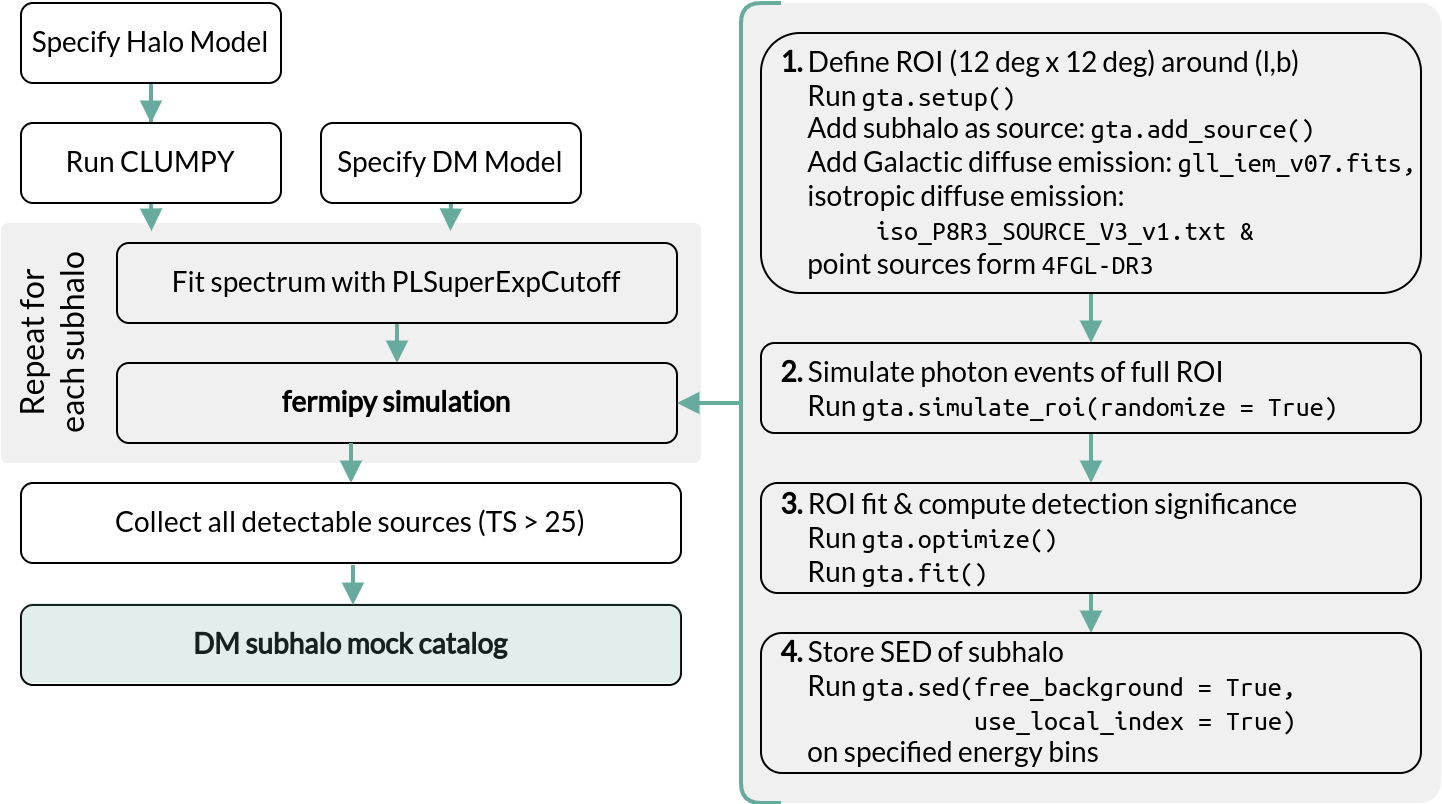}
		\caption{Flowchart of the subhalo simulation pipeline with detailed overview of the used \textsc{fermipy} subroutines (see also sections~\ref{sec:data_sim} and~\ref{app:fermipy_details}). The pipeline shown here is applied to each dark matter model, defined by ($\mDM, \sigv$), and the fixed annihilation channel.}
		\label{fig:flowchart}
\end{figure}

The \textsc{fermipy}~high-level analysis of \emph{Fermi}-LAT data permits to generate simulated data, analyse them to obtain the spectral energy distribution and the significance of point-like and extended sources.  
We employ \textsc{fermipy}~version~v1.0.1  and the Fermitools version~2.0.8. 
The analysis is performed by creating an instance of GTAnalysis (referred to as \texttt{gta} in what follows), 
which acts as a wrapper over the methods implemented in the Fermi Science Tools to fix or free parameters, add or remove sources from the model, and perform a fit to the ROI. 
Below, we supplement the description of the simulation provided in the main text with further technical details for each step to facilitate the reproducibility of our results.  
Our subhalo simulation pipeline is also summarised in the flowchart in figure~\ref{fig:flowchart} and table~\ref{tab:fermipy_stats}. 

\begin{enumerate}
\item  \textit{Setup.}
    After the initial setup of the ROI (\texttt{gta.setup()}), we add the subhalo source to the ROI model (\texttt{gta.add\_source()}), using as the spectral model a power law with exponential cutoff (PLSuperExpCutoff) with the parameters derived as in section~\ref{sec:data_sim}.  The Galactic diffuse emission is taken from the model optimised for the 4FGL-DR3 catalog (\texttt{gll\_iem\_v07.fits}), together with the corresponding isotropic diffuse emission model (\texttt{iso\_P8R3\_SOURCE\_V3\_v1.txt}), while the point-like and extended sources are taken from the 4FGL-DR3 catalog. All sources are inserted into the model with freely variable parameters (normalisation, spectral index). Only a single dark matter subhalo is considered for each ROI. In a more realistic situation, several subhalos would contribute to the $\gamma$-rays observed in a 12 x 12 degree region of the sky. In this case, further systematic uncertainties would arise from fitting their spectral parameters.  We have verified that the angular separation between bright subhalos is large enough to justify our approximation. 
\item  \textit{Photon event simulation.}
    We run \texttt{gta.simulate\_roi()} to simulate the ROI. We select the option \texttt{randomize= True} to randomize the data using Poisson statistics. This is done to include possible effects due to imperfect knowledge of the background in the simulation of the subhalo spectrum, see also the discussion in ref.~\cite{DiMauro:2020uos}.      
\item \textit{ROI fit.} 
    We proceed with the simulation reconstruction using the \texttt{gta.optimize()} and \texttt{gta.fit()} routines iteratively until a good fit quality (\texttt{fit\_quality} = 3) is reached. At the end of the fit, the TS of the central dark matter subhalo in the full energy range is obtained and stored. 
\item \textit{Spectral energy distribution.} 
   We use \texttt{gta.sed()} to obtain the spectral energy distribution (SED) of the simulated dark matter subhalo. We run  \texttt{gta.sed()} with the options \texttt{free\_background=True} to profile the normalisation of all background components in the model, and set \texttt{use\_local\_index = True} to use the correct power-law approximation to the shape of the global spectrum in each bin.  
\end{enumerate}

\begin{table*}[]
\def\arraystretch{1.15}
\caption{Summary of the \textsc{fermipy} analysis setup.}
\center
\begin{tabular}{ll}
\toprule
Time domain (MET) & 239557417 to 620181124 \\
Energy range & 500 MeV - 1 TeV \\
IRFs & \texttt{P8R3\_SOURCE\_V3} \\
Event type & FRONT+BACK \\
Point-source catalog &  4FGL–DR3 \\
ROI size & $12^{\circ} \times 12^{\circ}$ \\
Max zenith angle & $105^{\circ}$ \\
Galactic diffuse model & \texttt{gll\_iem\_v07.fits} \\
Isotropic diffuse model & \texttt{iso\_P8R3\_SOURCE\_V3\_v1.txt} \\
\bottomrule
\end{tabular}
 \label{tab:fermipy_stats}
\end{table*}

\subsection{Bayesian neural network optimisation}
\label{app:architecture}
Bayesian neural networks (BNNs) are a type of neural networks that replace the individual weights of a deep neural network (DNN) with weight distributions. This allows the evaluation of a distribution over likely output values for a fixed input, and the assignment of an uncertainty to the prediction. During training, BNNs learn the true model posterior, which is the distribution over network weights given a dataset. However, due to the complexity of the model, the true posterior cannot be directly inferred. Instead, the posterior is approximated by variational weight distributions with learnable parameters, and the difference between the estimated posterior and the prior is minimised by varying the parameters. The Kullback-Leibler divergence is used as the difference measure, and its average over the training dataset is used to compute the regularisation of the BNN. 

In this work, we implemented the network in the \textsc{Tensorflow} API using the \textsc{Tensorflow probability} framework. The network consists of multiple hidden layers, each of which is a \texttt{DenseFlipout} layer, following the algorithm proposed in \cite{flipout}. The flipout algorithm is efficient in this type of setup because it uses a so-called reparameterisation trick to maintain the independence of weight updates while still allowing for efficient gradient computation. This allows the gradients to be computed using the standard backpropagation algorithm, without the need to resample the weights.

The architecture of the network is defined by the type of layers (see previous paragraph) and its hyperparameters, such as the width and depth of the network and its activation function. Since there is no ideal set of hyperparameters to create the optimal network, and the final network performance will always depend on the weight initialisations and the efficiency of the gradient descent algorithm, the parameters can be optimised to some extent. With a good intuition of an appropriate architecture from previous work \cite{Butter:2021mwl}, we performed an architecture search, explicitly testing different reasonable options to find the most appropriate choice. This approach is typically not time efficient, but is sufficient here because both the network and the training sets are small, leading to short training times. In addition, this approach has the advantage that we can gain a more thorough understanding of the network's behaviour.  During the optimisation process, we learned that the performance of the network is very robust to changes in the architecture, i.e., the number of hidden layers. The final set of hyperparameters we chose for our networks is listed in table~\ref{tab:param}, with the number of hidden layers being the largest among the tested working configurations. 
 
\begin{table*}[]
\def\arraystretch{1.15}
\caption{Hyperparameter settings for the BNN architecture used in this work, see section~\ref{sec:ml_class}.}
\center
\begin{tabular}{ll}
\toprule
Weight prior & $\mathcal{N} (0, 2)$ \\
Kernel divergence & KL-Divergence \\
Number of hidden layers        &    4   \\
Number of nodes per layer          &    16  \\
Batch size     &    128  \\
Number of training epochs  &    200 + \texttt{EarlyStopping} (patience = 25) \\
\multirow{2}{*}{Learning rate of Adam optimizer} &  $10^{-3}$ \\ & + \texttt{ReduceLROnPlateau}   (patience = 20, min = $10^{-5}$)\\ 
\bottomrule
\end{tabular}
 \label{tab:param}
\end{table*}

\section{Supplementary results for 1 TeV}\label{app:1tev}
\begin{figure}[t]
	\center
	\includegraphics[width=.65\textwidth]{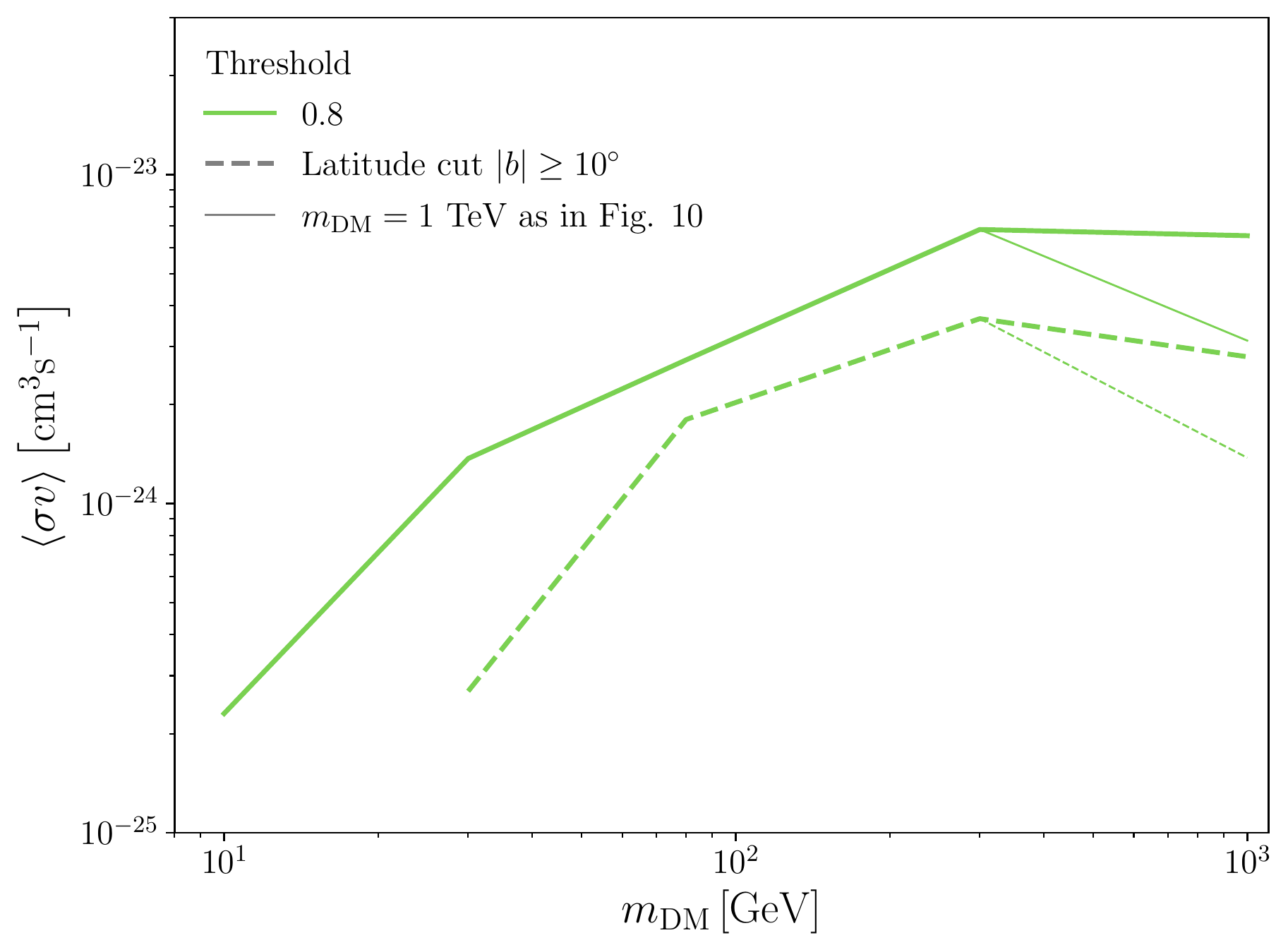}
		\caption{Limits on the annihilation cross section as in figure~\ref{fig:sigv_limits_mDM} for the $t = 0.8$ classification threshold. Here we show the effect on the limits for $m_{\rm DM}=1$~TeV when excluding the high energy bins as discussed in section~\ref{sec:results_det} \textit{(thick lines)}, compared to the standard treatment \textit{(thin lines)}. Results are shown without (solid lines) and with (dashed lines) the latitude cut.}
		\label{fig:sigv_limits_mDM_1TeV}
\end{figure}

As explained in section~\ref{sec:results_det}, we have verified that the number of detectable subhalos for $m_{\rm DM}=1$~TeV is sensitive to the energy bins considered to compute the TS. In particular, when considering the energy bins above 30~GeV, we are more sensitive to high dark matter masses, as the spectra peak in this range. We illustrate in figure~\ref{fig:sigv_limits_mDM_1TeV} the effect on the annihilation cross section limits of considering only the energy spectrum up to 30~GeV in the detectability analysis (thick lines), compared to the treatment presented in the main text (thin lines). Independent of the latitude cut, we find that ignoring the energy spectrum above 30~GeV would decrease the limits by a factor of two. 

\bibliography{astroml_ep3}
\bibliographystyle{JHEP}

\end{document}